\documentclass{Setting/SciPost}

\hypersetup{
    colorlinks,
    linkcolor={red!50!black},
    citecolor={blue!50!black},
    urlcolor={blue!80!black}
}

\usepackage[bitstream-charter]{mathdesign}
\DeclareSymbolFont{usualmathcal}{OMS}{cmsy}{m}{n}
\DeclareSymbolFontAlphabet{\mathcal}{usualmathcal}

\fancypagestyle{SPstyle}{
\fancyhf{}
\lhead{\colorbox{scipostblue}{\bf \color{white} ~SciPost Physics }}
\rhead{{\bf \color{scipostdeepblue} ~Submission }}

\fancyfoot[C]{\textbf{\thepage}}
}

\usepackage{bm}                             
\usepackage{slashed}                        
\usepackage{physics}                        
\usepackage{mathtools}                      

\usepackage{epstopdf}   
\usepackage{subcaption} 
\usepackage{float}      

\usepackage{array}
\usepackage{booktabs}    
\usepackage{multirow}
\usepackage{tabularx}    

\usepackage{pifont}      

\usepackage{tikz}        
\usetikzlibrary{positioning}            
\usetikzlibrary{shapes.geometric}       
\usetikzlibrary{arrows.meta}
\usetikzlibrary{decorations.pathmorphing} 

\usepackage{rotating}    
\usepackage{pgfplots}    
\pgfplotsset{compat=1.18} 
\usepackage{pgf-pie}     

\usepackage[compat=1.1.0]{tikz-feynman}
\tikzfeynmanset{warn luatex=false}

\let\oldepsilon\epsilon       
\let\oldvarepsilon\varepsilon 

\renewcommand{\epsilon}{\oldvarepsilon}
\renewcommand{\varepsilon}{\oldepsilon}

\newcommand{\cmark}{\ding{51}}

\usepackage{tikz,pgfplots}

\usetikzlibrary{positioning} 
\usetikzlibrary{arrows.meta} 
\usetikzlibrary{calc} 

\colorlet{myred}{red!80!black}
\colorlet{myblue}{blue!80!black}
\colorlet{mygreen}{green!50!black}
\colorlet{myorange}{orange!80!yellow!90!red!90!black}

\tikzstyle{mycomment}=[inner sep=1pt,scale=0.75,align=left]

\tikzstyle{mybox}=[draw,#1!80!black,fill=#1!95!black!20,inner sep=5pt,outer sep=3pt,
                   thick,rounded corners=3pt,align=center,font=\bfseries]

\tikzstyle{myarrow}=[-{Latex[length=8,width=8]},#1!80!black,thick,line cap=round,line width=3]

\def\connect[#1](#2)!#3!(#4){
  \draw[#1] (#2) |- ($(#2)!#3!(#4)$) node[pos=0.5] (#2-#4-1) {}
  -| (#4) node[pos=0.5] (#2-#4-2) {}
}

\begin{document}

\pagestyle{SPstyle}

\begin{center}{\Large \textbf{\color{scipostdeepblue}{
On the efficiency of parameter space exploration:\\ A scotogenic case study\\
}}}\end{center}

\begin{center}\textbf{
Ugo de Noyers\textsuperscript{1$\star$},
Mathis Dubau\textsuperscript{2$\dagger$},
Bj\"orn Herrmann\textsuperscript{1$\ddagger$} and
Olivier Arnaez\textsuperscript{2$\circ$}
}\end{center}

\begin{center}
{\bf 1} LAPTh, Univ.\ Savoie Mont Blanc, CNRS, F-74000 Annecy, France
\\
{\bf 2} LAPP, Univ.\ Savoie Mont Blanc, CNRS, F-74000 Annecy, France
\\[\baselineskip]
$\star$ \href{mailto:denoyers@lapth.cnrs.fr}{\small denoyers@lapth.cnrs.fr}\,,\quad
$\dagger$ \href{mailto:mathis.dubau@cern.ch}{\small mathis.dubau@cern.ch}\,,\quad
$\ddagger$ \href{mailto:herrmann@lapth.cnrs.fr}{\small herrmann@lapth.cnrs.fr}\,,\\
$\circ$ \href{mailto:arnaez@lapp.in2p3.fr}{\small arnaez@lapp.in2p3.fr} 
\end{center}

\section*{\color{scipostdeepblue}{Abstract}}
\textbf{\boldmath{%
A common problem in beyond Standard Model phenomenology is the exploration of a multi-dimensional parameter space in view of a large number of constraints. We study and compare two methods applicable to this challenge, namely a Markov Chain Monte Carlo scan (MCMC) and a Deep Neural Network (DNN). We illustrate both methods via their application to different scotogenic frameworks, allowing to extend the Standard Model to include viable dark matter candidates while generating neutrino mass terms at the one-loop level. Our studies allow us to compare the two employed methods, both at the level of phenomenology and at the level of computing effort. We find that, while phenomenologically speaking both methods deliver compatible conclusions, the obtained datasets feature differences at the detail level in the distributions of observables, e.g.\ the dark matter mass.
}}

\vspace{\baselineskip}

\noindent\textcolor{white!90!black}{%
\fbox{\parbox{0.975\linewidth}{%
\textcolor{white!40!black}{\begin{tabular}{lr}%
  \begin{minipage}{0.6\textwidth}%
    {\small Copyright attribution to authors. \newline
    This work is a submission to SciPost Physics. \newline
    License information to appear upon publication. \newline
    Publication information to appear upon publication.}
  \end{minipage} & \begin{minipage}{0.4\textwidth}
    {\small Received Date \newline Accepted Date \newline Published Date}%
  \end{minipage}
\end{tabular}}
}}
}


\vspace{10pt}
\noindent\rule{\textwidth}{1pt}
\tableofcontents
\noindent\rule{\textwidth}{1pt}
\vspace{10pt}

\section{Introduction}
\label{Sec:Intro}

The Standard Model (SM) of particle physics is the prevailing theory that describes three of the fundamental interactions, namely the strong, weak, and electromagnetic forces, and classifies all the observed particles into bosons and fermions. It has successfully predicted a wide range of phenomena and its predictions have been tested to very good precision in numerous experiments over several energy scales. Despite these successes, the SM has notable limitations and leaves open questions. In addition to shortcomings of theoretical nature, it lacks a candidate for dark matter observed in our Universe \cite{Planck2018}, and predicts massless neutrinos in contradiction with experimental evidence of neutrino oscillations \cite{NuFit2020}.

In order to address these open questions, the SM needs to be extended, and theories beyond the SM (BSM) are explored. A first key area of investigation is the origin of neutrino masses. In the SM, unlike other fermions, neutrinos do not acquire mass through the Higgs mechanism. The true nature of neutrinos—whether they are Dirac or Majorana particles—remains unresolved, allowing for different mass generation mechanisms in BSM models. One possibility is the Seesaw mechanism \cite{Minkowski1977, Yanagida1980, PhysRevLett.56.561}, including heavy ``right-handed'' neutrinos. In the present work, we consider the possibility of Majorana neutrinos acquiring mass terms induced at the one-loop level \cite{Ma2006}.

A second indication for BSM physics is the presence of dark matter (DM) in our Universe. Astrophysical and cosmological observations strongly suggest that dark matter accounts for about $27\%$ of the Universe’s energy density, yet it remains undetected in laboratory experiments. A promising dark matter candidate is the Weakly Interacting Massive Particle (WIMP), which is massive, electrically neutral, and stable. WIMPs can successfully account for the observed relic density of cold dark matter (CDM) through a process known as thermal freeze-out \cite{Planck2018}.

In this paper, we focus on a framework that addresses both the generation of neutrino masses and the nature of dark matter. The two aspects are linked by the fact that the neutrino masses are induced through loop diagrams containing states from the dark sector, leading to what is known as a ``scotogenic'' model, originally proposed in Ref.\ \cite{Ma2006}. Depending on the exact field content, such scotogenic scenarios may be realized based on different topologies and associated field contents for radiative neutrino mass generation \cite{Yaguna2013}. This class of models has gained considerable attention over the last decade \cite{Toma:2013zsa, Vicente:2014wga, Fraser:2014yha, Betancur:2017dhy, Betancur:2018xtj, Baumholzer:2019twf, Esch2018, Sarazin:2021nwo, Alvarez:2023dzz, deNoyers:2024qjz, Darricau:2025vcs} due to their ability to simultaneously address two observationally motivated open questions of the SM of particle physics. 

An important consequence of the chosen Majorana mass generation mechanism is the presence of lepton-flavour violating effects, occurring at energy scales that are reachable at current experiments. Such transitions are forbidden in the SM and experimentally strongly constrained \cite{ParticleDataGroup:2022pth}, especially in the case of $\mu-e$ transitions. Consequently, they put strong restrictions on the physically viable parameter space of the considered framework. Additional constraints stem from the observed mass of the Higgs boson \cite{ParticleDataGroup:2022pth} and the cold dark matter relic density, which is determined to very high precision in the $\Lambda$CDM cosmological model \cite{Planck2018}.

In the present study, for all scotogenic frameworks under consideration, we assume the presence of an additional $\mathbb{Z}_2$ discrete symmetry, allowing us to ensure stability of the dark matter candidate as per cosmological constraints. Generally, scotogenic frameworks provide three possible viable dark matter candidates, while generating at least two non-zero neutrino masses, allowing to meet the constraints on the observed neutrino mass differences. $CP$-violating is encompassed through the Dirac as well as two potential Majorana phases in the Pontecorvo-Maki-Nakagawa-Sakata (PMNS) matrix \cite{Pontecorvo1957b, Maki1962}. 

The goals of the present analysis are two-fold: One the one hand, we intend to investigate the phenomenological aspects of given scotogenic realizations. On the other hand, we intend to compare two different methods of parameter space exploration, the first being a Markov Chain Monte Carlo (MCMC) scan, the second being a scan using a Deep Neural Network (DNN). The MCMC technique has already been used in previous studies by some of us \cite{Sarazin:2021nwo, Alvarez:2023dzz, deNoyers:2024qjz}, while the DNN technique has, to the best of our knowledge, not been employed in similar parameter space studies. In this spirit, the present analysis is complementary to those published in Refs.\ \cite{Sarazin:2021nwo, Alvarez:2023dzz, deNoyers:2024qjz}, allowing to confront the two scanning techniques on both the phenomenological and computational level. In practice, we choose to work within two scotogenic frameworks: The first, labelled ``T1-2G'' according to Ref.\ \cite{Yaguna2013}, has already been studied in Ref.\ \cite{deNoyers:2024qjz} using the MCMC scanning technique. It therefore can serve as benchmark for the comparison with the newly employed DNN scanning technique. The second framework under consideration is labelled ``T1-2B'' and is complementary to the previous studies of Ref.\ \cite{Sarazin:2021nwo, Alvarez:2023dzz, deNoyers:2024qjz} through the presence of scalar triplet. 

The present paper is organised as follows: We present the two considered models in Sec.\ \ref{Sec:Models}. Sec.\ \ref{Sec:Setup} is devoted to the discussion of the relevant constraints and the two computational setups to be employed. Limiting ourselves to the case of fermionic dark matter, we discuss our results on the phenomenological level in Sec.\ \ref{Sec:Results} and on the computational level in Sec.\ \ref{Sec:Comparison}, before concluding in Sec.\ \ref{Sec:Conclusion}. Additional technical details can be found in the Appendices. Note that, for the sake of a concise presentation and in order to be able to focus on the novel aspects of this work, for certain aspects of the phenomenological discussion -- e.g.\ aspects of the mass spectrum or previously studied frameworks -- the reader will simply be referred to previous publications, mainly Refs.\ \cite{Sarazin:2021nwo, Alvarez:2023dzz, deNoyers:2024qjz}.

\section{Scotogenic models under consideration}
\label{Sec:Models}

In the present study, we will consider two scotogenic frameworks, featuring different field contents and therefore differing in certain phenomenological aspects. In this section, we will briefly introduce the two frameworks and discuss their similarities and differences.

\subsection{Field content and mass eigenstates}
\label{Sec:FieldsMassEigenstates}

The first scotogenic framework to be considered in this work comprises a scalar singlet, a scalar doublet, a fermionic Dirac doublet, and two generations of a fermionic triplet, as summarized in the last row of Table \ref{Tab:GaugeEigenstates}. This configuration is labelled ``T1-2G'' in the classification of Ref.\ \cite{Yaguna2013}. Thanks to the presence of two triplets, three non-zero neutrino masses are generated. A phenomenological discussion of this model, based solely on an MCMC analysis, can be found in Ref.\ \cite{deNoyers:2024qjz}.

The second framework under consideration is based on the model labelled ``T1-2B'' according to the classification of Ref.\ \cite{Yaguna2013}. It comprises two fermionic singlets, a fermionic Dirac doublet, a complex scalar doublet and a complex scalar triplet. This setup shares the fermionic sector of the ``T1-2A'' model \cite{Sarazin:2021nwo, Alvarez:2023dzz}, but differs in the scalar sector due to the presence of the triplet. Note that the initial ``T1-2B'' model features only one fermionic singlet. This configuration allows for the generation of two non-zero neutrino masses, which is per se sufficient to satisfy the constraints from neutrino oscillations. However, an additional degree of freedom is required to achieve a third non-zero neutrino mass, which may appear somewhat more natural. The simplest way to realize this is adding an additional fermionic singlet, which leads to the model labelled as ``T1-2B ext.'' (extended) throughout the present paper. The field content of both frameworks is listed in the upper part of Table \ref{Tab:GaugeEigenstates}.

\begin{table}
    \centering
    \begin{tabular}{|c||c|c|c|c|c|c||c|c|c|}
    \hline
    & ~$F_1$~ & ~$F_2$~ & ~$\Psi_1$~ & ~$\Psi_2$~ & ~$\Sigma_1$~ & ~$\Sigma_2$~ & ~$S$~ & ~$\eta$~ & ~$\Delta$~  \\
    \hline\hline
    $SU(2)_{L}$ & $\mathbf{1}$ & $\mathbf{1}$ & $\mathbf{2}$ & $\mathbf{2}$ & $\mathbf{3}$ & $\mathbf{3}$ & $\mathbf{1}$ & $\mathbf{2}$ & $\mathbf{3}$ \\
    $U(1)_{Y}$   & 0 & 0 & -1 & -1 & 0 & 0 & 0 & 1 & 0 \\
    \hline
    \hline
    ``T1-2B''       & \cmark & -- & \cmark & \cmark & -- & -- & \cmark & -- & \cmark       \\
    ``T1-2B ext.''       & \cmark & \cmark & \cmark & \cmark & -- & -- & \cmark & -- & \cmark       \\
    \hline
    ``T1-2G''       & -- & -- & \cmark & \cmark & \cmark & \cmark & \cmark & \cmark & --       \\
    \hline
    \end{tabular}
    \caption{Summary of the additional fields in the scotogenic models ``T1-2B'', ``T1-2B ext.'', and ``T1-2G''.}
    \label{Tab:GaugeEigenstates}
\end{table}

Using the field notation introduced in Table \ref{Tab:GaugeEigenstates}, the Lagrangians of the two presented frameworks include the following terms for the scalar potential,
\begin{align}
    -{\cal L}^{\rm ``T1-2B~ext."} ~&\supset~ M^2_H |H|^2 + \frac{1}{2} M^2_{\Delta} \text{Tr}\big\{ \Delta^{\dagger} \Delta \big\} + \lambda_H |H|^4 + M^2_{\eta} |\eta|^2 + \lambda_{4\eta} |\eta|^4 \nonumber \\
        &~~~~~ + \frac{1}{2} \lambda_{\Delta} |H^{\dagger} \Delta|^2 + \frac{1}{2} \lambda_{\Delta \eta} |\eta^{\dagger} \Delta|^2 + \frac{3}{2} \lambda_{4\Delta} \text{Tr}\big\{( \Delta^{\dagger} \Delta )^2 \big\} \\ 
        &~~~~~ + \lambda_{\eta} |\eta|^2 |H|^2 + \lambda^{\prime}_{\eta} |\eta^{\dagger} H|^2 + \frac{1}{2} \lambda^{\prime\prime}_{\eta} \big( ( \eta^{\dagger} H )^2 + \text{h.c.} \big) + \kappa \big( \eta^{\dagger} \Delta H + \text{h.c.} \big) \,, \nonumber \\ 
    -{\cal L}^{\rm ``T1-2G"} ~&\supset~ M^2_H \vert H \vert^2 + \lambda_H \vert H \vert^4 
        + \frac{1}{2} M^2_S S^2 + \frac{1}{2} \lambda_{4S} S^4 + M^2_{\eta} \vert \eta \vert^2 + \lambda_{4\eta} \vert \eta \vert^4 \nonumber \\ 
        &~~~~~ + \frac{1}{2} \lambda_S S^2 \vert H \vert^2  + \frac{1}{2} \lambda_{S \eta} S^2 \vert \eta \vert^2 + \lambda_{\eta} \vert \eta \vert^2 \vert H \vert^2 +  \lambda^{\prime}_{\eta} \vert \eta^{\dagger} H \vert^2 \\ 
        &~~~~~ + \frac{1}{2} \lambda^{\prime\prime}_{\eta} \left( \big( \eta^{\dagger} H  \big)^2 + \text{h.c.} \right) + \kappa \left( S \eta^{\dagger} H + \text{h.c.} \right) \,, \nonumber
\end{align}
and the following terms leading to fermion masses,
\begin{align}
    -{\cal L}^{\rm ``T1-2B~ext."} ~&\supset~ M_{\Psi} \Psi_1 \Psi_2 + \frac{1}{2} M_{F_{ij}} \bar{F}_i F_j + y_1^j \Psi_1 F_j H + y_2^j \Psi_2 F_j \tilde{H} + \text{h.c.} \,, \\
    -{\cal L}^{\rm ``T1-2G"} ~&\supset~ M_{\Psi} \Psi_1 \Psi_2 + \frac{1}{2} M_{\Sigma}^{ij} \, \mathrm{Tr}\!\left\{ \overline{\Sigma}_i \Sigma_j  \right\} + y_1^j \Psi_1 \Sigma_j H + y_2^j \Psi_2 \Sigma_j \tilde{H} + \text{h.c.} \,, 
\end{align}
where $\tilde{H} = i \sigma_2 H^c$.

After electroweak symmetry breaking, the components of the fields displayed in Table \ref{Tab:GaugeEigenstates}, that share the same quantum numbers, will mix to give rise to a series of physical mass eigenstates (see App.\ \ref{App:T12B} for details). In each of the above models, this mixing will result in a number of physical neutral fermions, charged fermions, neutral scalars, charged scalars, and pseudo-scalars. For each model, the respective numbers are summarized in Table \ref{Tab:MassEigenstates}. Depending on the exact mass hierarchy, the dark matter candidate can either be the lightest neutral fermion, $\chi^0_1$, the lightest neutral scalar, $\phi^0_1$, or the pseudo-scalar, $A^0$. In the subsequent analysis, we limit ourselves to the case of fermionic dark matter.

\begin{table}
    \centering
    \begin{tabular}{|c||cc||cc|c|}
        \hline
        Model & \multicolumn{2}{c||}{Fermions} & \multicolumn{2}{c|}{Scalars} & Pseudo-scalars \\
              & neutral & charged & neutral & charged &   \\
        \hline
        \hline
         ``T1-2B'' & $\chi^0_1$, $\chi^0_2$, $\chi^0_3$ & $\chi^{\pm}$ & $\phi^0_1$, $\phi^0_2$ &  $\phi^{\pm}_1$, $\phi^{\pm}_2$ & $A^0$ \\
         ``T1-2B ext.'' & $\chi^0_1$, $\chi^0_2$, $\chi^0_3$, $\chi^0_4$ & $\chi^{\pm}$ & $\phi^0_1$, $\phi^0_2$ & $\phi^{\pm}_1$, $\phi^{\pm}_2$ & $A^0$ \\
         \hline
         ``T1-2G'' & $\chi^0_1$, $\chi^0_2$, $\chi^0_3$, $\chi^0_4$ & $\chi^{\pm}_1$, $\chi^{\pm}_2$, $\chi^{\pm}_3$ & $\phi^0_1$, $\phi^0_2$ & $\phi^{\pm}$ & $A^0$ \\
        \hline         
    \end{tabular}
    \caption{Physical mass eigenstates of the scotogenic frameworks defined in Table \ref{Tab:GaugeEigenstates}.}
    \label{Tab:MassEigenstates}
\end{table}

\subsection{Interaction terms}
\label{Sec:InteractionTerms}

A key ingredient of scotogenic models is the interaction terms between the additional scalar, additional fermionic fields, and the Standard Model lepton doublets. In addition to their contribution to neutrino mass generation, these terms induce lepton flavour violating interactions at the one-loop level, which can be constrained by experimental measurements. 

For the three models under consideration, the respective interaction Lagrangians read
\begin{align}
    - {\cal L}^{\rm ``T1-2B~ext."} ~&\supset~ g_{\Psi}^{\alpha} \Psi_2 \Delta L_{\alpha} + g_{F_j}^{\alpha} F_j \eta L_{\alpha} + g_R^{\alpha} \ell_{\alpha}^c \tilde{\eta} \Psi_1 + \text{h.c.} \,, \\
    - {\cal L}^{\rm ``T1-2G''} ~&\supset~ g_{\Psi}^{\alpha} L_{\alpha} \Psi_2 S + g_{\Sigma_j}^{\alpha} \eta \Sigma_j L_{\alpha}  + g_{R}^{\alpha} \ell_{\alpha}^c \tilde{\eta} \Psi_1 + \text{h.c.} \,.
\end{align}
Here, $L$ denotes the Standard Model lepton doublet, and the index $\alpha$ runs over the three generations of leptons ($\alpha = e, \mu, \tau$). The notation $\tilde{\eta}$ corresponds to $\tilde{\eta} = i \sigma_2 \eta^c$, and sums over the indices $j$ and $\alpha$ are implicit.

The Yukawa-type couplings appearing in the Lagrangians above can conveniently be regrouped in a $3\times 3$ matrix
\begin{equation}
    {\cal G} ~=~ \begin{pmatrix} g_{\Psi}^e & g_{\Psi}^e & g_{\Psi}^{\tau} \\ g_{1}^e & g_{1}^{\mu} & g_{1}^{\tau} \\ g_{2}^e & g_{2}^{\mu} & g_{2}^{\tau} \end{pmatrix} \,,
\end{equation}
where $g_i^{e,\mu,\tau} = g_{F_i}^{e,\mu,\tau}$ ($i=1,2$) for the case of ``T1-2B ext.'', while $g_i^{e,\mu,\tau} = g_{\Sigma_i}^{e,\mu,\tau}$ ($i=1,2$) for the case of ``T1-2G''.

\subsection{Neutrino mass generation}
\label{Sec:NeutrinoMassGeneration}

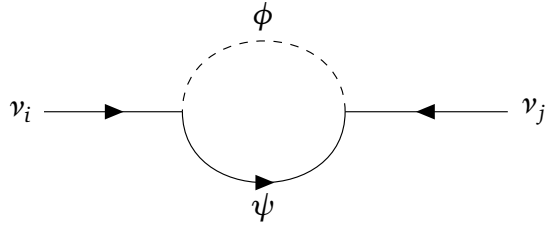
\begin{figure}
    \centering
    \resizebox{0.5\columnwidth}{!}{%
      \begin{tikzpicture}
            \begin{feynman}
                \vertex (start) at (0,0) {\( \nu_i \)};
                \vertex[right=2cm of start] (loopstart);
                \vertex[right=2cm of loopstart] (loopend);
                \vertex[right=2cm of loopend] (end) {\( \nu_j \)};
                \diagram* {
                    (start) -- [fermion] (loopstart),
                    (loopstart) -- [fermion, half right, edge label'=\(\psi\)] (loopend),
                    (loopend) -- [scalar,  half right, edge label'=\(\phi\)] (loopstart),
                    (end)   -- [fermion] (loopend),
                };
            \end{feynman}
        \end{tikzpicture}
    }
    \caption{Radiative generation of neutrino masses within the a scotogenic model, depicted in the interaction basis. Here, $\psi$ is a generic notation for a fermionic field (singlet $F$, doublet $\Psi$, or triplet $\Sigma$), while $\phi$ is a generic notation for a scalar field (singlet $S$, doublet $\eta$, or triplet $\Delta$).}
    \label{Fig:NeutrinoMassGeneration}
\end{figure}

In each considered scotogenic model, neutrino masses are generated via one-loop diagrams as shown in Fig.\ \ref{Fig:NeutrinoMassGeneration}, where the additional scalar and fermionic degrees of freedom appear in the loop. The corresponding mass term in the Lagrangian can be expressed as
\begin{equation}
    {\cal L} ~\supset~ -\frac{1}{2} \, \bar{\nu}^c \, {\cal G}^t \, M_L \, {\cal G} \, \nu \, ,
    \label{Eq:NeutrinoMassMatrix}
\end{equation}
where the couplings are factorized, and the matrix $M_L$ contains the actual expressions of the loop integrals appearing in Fig.\ \ref{Fig:NeutrinoMassGeneration} depending on the parameters related to the respective scalar and fermionic sectors. Explicit expressions for the elements of $M_L$ can be found in Ref.\ \cite{deNoyers:2024qjz} for ``T1-2G'', and App.\ \ref{App:T12B} for ``T1-2B'' and ``T1-2B ext.''.

In practice, following the Casas-Ibarra parametrization \cite{CasasIbarra2001}, the coupling matrix ${\cal G}$ is computed from the scalar and fermion mass parameters, combined with the actual neutrino mass data, according to
\begin{equation}
    {\cal G} ~=~ U_L \, D_L^{-1/2} \, R \, D_{\nu}^{1/2} \, U^{*}_{\rm PMNS} \,.
    \label{Eq:CasasIbarra}
\end{equation}
Here, $U_{\rm PMNS}$ is the PMNS-matrix encompassing information about neutrino mixing and $CP$-violating phases, $D_{\nu}$ is the diagonal matrix comprising the neutrino mass eigenvalues, $D_L$ and $U_L$ are related to the matrix $M_L$ such that $D_L ~=~ U_L^{\dag} M_L U_L$, and $R$\ is a complex rotation matrix involving three complex parameters $r_k$ ($k=1,2,3$) as detailed in App. \ref{App:CasasIbarra}. This procedure allows us to efficiently choose the couplings conveniently in order to meet the constraints stemming from the neutrino sector. For a detailed discussion of the Casas-Ibarra parametrization in scotogenic models see, e.g., Refs.\ \cite{Sarazin:2021nwo, deNoyers:2024qjz}.

\subsection{Physical mass spectrum}
\label{Sec:PhysicalMassSpectrum}

As mentioned above, the physical mass eigenstates are obtained by diagonalizing the respective mass matrices appearing in the Lagrangian of the exact framework under consideration. While the Lagrangian allows to extract the mass matrices at the tree-level, one-loop corrections to the masses can be of phenomenological importance \cite{Sarazin:2021nwo, deNoyers:2024qjz}.

Throughout our analyses, we make use of the numerical spectrum calculator {\tt SPheno} \cite{SPheno2003, SPheno2011}, which allows to compute the mass eigenvalues and associated mixing matrices of a new physics model including one-loop corrections. To this end, we have implemented the scotogenic models discussed above using the {\tt Mathematica} package {\tt SARAH} \cite{SARAH2008, SARAH2010, SARAH2011, SARAH2013, SARAH2014}. The latter generated the numerical modules for {\tt SPHENO}.
\section{Computational setup and tools}
\label{Sec:Setup}

In this Section, we present the various setups and computational tools used to perform the analyses presented in this study. We will also give the studied parameter ranges as well as the experimental constraints considered in the parameter space exploration.

Being interested in the phenomenology of the scotogenic frameworks introduced above, we impose a number of experimental constraints on the multi-dimensional parameter space in order to determine the viable parameter regions. As already mentioned in the Introduction, we do so by employing two different numerical techniques, namely the Markov Chain Monte Carlo (MCMC) scanning technique and a Deep Neural Network (DNN) technique. 

In Table \ref{Tab:ScanParameters}, we summarize the model parameters associated to each considered framework, ``T1-2B ext.'' and ``T1-2G'', together with the respective intervals imposed on the algorithm. The imposed intervals are chosen so as to lead to interesting configurations with TeV-scale phenomenology, i.e.\ not having too high masses, while remaining perturbatively stable, i.e.\ not featuring too high coupling values.

Moreover, we indicate in Table \ref{Tab:NeutrinoParameters} the numerical intervals for the input parameters used in the neutrino sector, namely the lightest neutrino mass $m_{\nu_1}$, the squared mass differences $\Delta m^2_{21}$ and $\Delta m^2_{31}$, the three mixing angles appearing in the PMNS matrix ($\theta_{12}$, $\theta_{13}$, and $\theta_{23}$), the Dirac ($\delta_{CP}$) and Majorana ($\alpha_1$ and $\alpha_2$) CP-violating phases, as well as the free complex parameters ($r_k$, $k=1,2,3$) associated to the rotation matrix $R$ as detailed in App.\ \ref{App:CasasIbarra}. The intervals for mass differences, mixing angles, and Dirac phase respect the constraints obtained from neutrino oscillation observations \cite{NuFit2024}, while the lightest neutrino mass is chosen so as not to violate the bound on the sum of neutrino masses \cite{ParticleDataGroup:2022pth}. Note that the Majorana phases and the parameter $r_k$ ($k=1,2,3$) are not constraint by observation.

\begin{table}
    \centering
        \begin{tabular}{|c|c|}
            \hline
            \multicolumn{2}{|c|}{\textbf{``T1-2B ext.''}} \\
            \hline
            \hline
            \textbf{Parameter} & \textbf{Range} \\
            \hline\hline
            $\lambda_H$ & $[0.1, 0.4]$ \\
            \hline
            $\lambda_{\Delta}, \lambda_{\Delta \eta}$ & $[-1, 1]$ \\
            \hline
            $\lambda_{4\Delta}, \lambda_{4\eta}$ & $[10^{-7}, 1]$ \\
            \hline
            $\lambda_{\eta}, \lambda^{\prime}_{\eta}, \lambda^{\prime\prime}_{\eta}$ & $[-1, 1]$ \\
            \hline
            $\kappa$ & $[-10^3, 10^3] \; \mathrm{GeV}$ \\
            \hline
            $M^2_{\Delta}, M^2_{\eta}$ & $[2 \cdot 10^{5}, 16 \cdot 10^{6}] \; \mathrm{GeV}^2$ \\
            \hline
            $M_{\Psi}, M_{F_{jj}}$ & $[6 \cdot 10^{2}, 4 \cdot 10^{3}] \; \mathrm{GeV}$ \\
            \hline
            $y_{ij}$ & $[-1, 1]$ \\
            \hline
            $g_R^{\alpha}$ & $[-1, 1]$ \\
            \hline
        \end{tabular}
        ~~
        \begin{tabular}{|c|c|}
            \hline
            \multicolumn{2}{|c|}{\textbf{``T1-2G''}} \\
            \hline
            \hline
             \textbf{Parameter} & \textbf{Interval} \\
             \hline\hline
            $\lambda_H$ & $\left[0.1, 0.4 \right]$ \\
            \hline
            $\lambda_{S}, \lambda_{S \eta}$ & $\left[-1, 1 \right]$ \\
            \hline
            $\lambda_{4S}, \lambda_{4 \eta}$ & $\left[10^{-7}, 1 \right]$ \\
            \hline
            $\lambda_{\eta}, \lambda^{\prime}_{\eta}, \lambda^{\prime\prime}_{\eta}$ & $\left[-1, 1 \right]$ \\
            \hline
            $\kappa$ & $\left[-10^3, 10^3 \right] \; \mathrm{GeV}$ \\
            \hline
            $M^2_S, M^2_{\eta}$ & $\left[2 \cdot 10^{5}, 16 \cdot 10^{6} \right] \; \mathrm{GeV}^2$ \\
            \hline
            $M_{\Psi}, M_{\Sigma_{jj}}$ & $\left[6 \cdot 10^{2}, 4 \cdot 10^{3} \right] \; \mathrm{GeV}$ \\
            \hline
            $y_{ij}$ & $\left[-1, 1 \right]$ \\
            \hline
            $g_R^{\alpha} $ & $\left[-1, 1 \right]$ \\
            \hline
        \end{tabular}
    \caption{Parameters varied in MCMC and DNN studies and associated parameter ranges for the ``T1-2B ext.'' (left) and ``T1-2G'' (right) frameworks. In our MCMC analyses, all parameters are scanned in logarithmic scale. Indices run such that $i,j \in \{1,2\}$, $k \in \{1,2,3\}$, and $\alpha \in \{e, \mu, \tau\}$.}
    \label{Tab:ScanParameters}
\end{table}

\begin{table}
    \centering
        \begin{tabular}{|c|c|}
            \hline
             \textbf{Parameter} & \textbf{Interval} \\
             \hline\hline
            $m_{\nu_1}$ & $\left[10^{-32}, 10^{-10} \right] \; \mathrm{GeV}$ \\
            \hline
            $\Delta m^2_{21}$ & $\left[7.01, 7.82\right] \cdot 10^{-23} \; \mathrm{GeV}^2$ \\
            \hline
            $\Delta m^2_{31}$ & $\left[2.457, 2.567\right] \cdot 10^{-21} \; \mathrm{GeV}^2$ \\
            \hline
            $\mathfrak{Re}(r_k)$, $\mathfrak{Im}(r_k)$ & $\left[-1, 1 \right]$ \\
            \hline
            $\alpha_{1}$, $\alpha_2$ & $\left[0^\circ, 180^\circ \right]$ \\
            \hline
            $\delta_{CP}$ & $\left[147^\circ, 281^\circ \right]$ \\
            \hline        
            $\theta_{12}$ & $\left[31.97^\circ, 34.91^\circ \right]$ \\
            \hline
            $\theta_{13}$ & $\left[8.3^\circ, 8.76^\circ \right]$ \\
            \hline        
            $\theta_{23}$ & $\left[46.5^\circ, 51.2^\circ \right]$ \\
            \hline
        \end{tabular}
    \caption{Parameters of the neutrino sector involved in the Casas-
Ibarra parametrization. In our MCMC analyses, $m_{\nu_1}$ is scanned in logarithmic scale, while all other parameters are scanned in linear scale. The index $k$ runs over $k \in \{1,2,3 \}$. The given intervals are taken from Ref.\ \cite{NuFit2024}.}
    \label{Tab:NeutrinoParameters}
\end{table}

Coming to the constraints, we impose the experimental determinations of the Higgs boson mass, the cold dark matter relic density, as well as a number of upper limits on lepton-flavour violating branching ratios and conversion rates. Moreover, we take into account the electric dipole moments (EDM) for the leptons, and -- for the ``T1-2G'' framework -- the anomalous magnetic moments of electron and muon\footnote{Note that, during the process of writing the present paper, new results in the anomalous magnetic moment of the muon have been released \cite{Muong-2:2025xyk}. Consequently, the current status is such that there no longer a significant deviation as compared to the Standard Model prediction. The deviation taken into account in our scan is \emph{per se} not valid any more. However, this has no impact on the validity of the studies models themselves and their ability to address other shortcomings of the Standard Model.}. The exact intervals, imposed at the $5\sigma$ confidence level, and upper limits are summarized in Table \ref{Tab:Constraints}. Note that, for the Higgs-boson mass and the cold dark matter relic density, the imposed interval in the DNN study is enlarged as compared to the interval imposed in the MCMC study. Numerically, we compute the above observables using {\tt SPHENO} (numerical modules generated by {\tt SARAH}), except for the dark-matter related observables, which are computed using {\tt micrOMEGAs} \cite{MO2001, MO2004, MO2007a, MO2007b, MO2013, MO2018}, where the necessary {\tt CalcHEP} \cite{CalcHep_Belyaev:2012qa} model files have been generated by {\tt SARAH}.

\begin{table}
    \centering
    \begin{minipage}{0.48\linewidth}
        \centering
        \begin{tabular}{|c|c|}
            \hline
            \textbf{Observable} & \textbf{Constraint} \\
            \hline\hline
            $m_H$ [GeV]& $125.25 \pm 3.0 ~(15.0)$ \\
            \hline
            $\Omega_{\mathrm{CDM}}h^2$ & \!\!$0.1200 \pm 0.0042 ~(0.0210)$ \!\!\!\! \\
            \hline
            \!$\mathrm{BR}(\mu^- \!\to\! e^- \gamma)$\! & $< 4.2 \cdot 10^{-13}$ \\
            \hline
            \!$\mathrm{BR}(\tau^- \!\to\! e^- \gamma)$\! & $< 3.3 \cdot 10^{-8}$ \\
            \hline
            \!$\mathrm{BR}(\tau^- \!\to\! \mu^- \gamma)$\! & $< 4.2 \cdot 10^{-8}$ \\
            \hline
            \!$\mathrm{BR}(\mu^- \!\to\! e^- e^+ e^-)$\! & $< 1.0 \cdot 10^{-12}$ \\
            \hline
            \!$\mathrm{BR}(\tau^- \!\to\! e^- e^+ e^-)$\! & $< 2.7 \cdot 10^{-8}$ \\
            \hline
            \!$\mathrm{BR}(\tau^- \!\to\! \mu^- \mu^+ \mu^-)$\! & $< 2.1 \cdot 10^{-8}$ \\
            \hline
            \!$\mathrm{BR}(\tau^- \!\to\! e^- \mu^+ \mu^-)$\! & $< 2.7 \cdot 10^{-8}$ \\
            \hline
            \!$\mathrm{BR}(\tau^- \!\to\! \mu^- e^+ e^-)$\! & $< 1.8 \cdot 10^{-8}$ \\
            \hline
            \!$\mathrm{BR}(\tau^- \!\to\! \mu^- e^+ \mu^-)$\! & $< 1.7 \cdot 10^{-8}$ \\
            \hline
            \!$\mathrm{BR}(\tau^- \!\to\! \mu^+ e^- e^-)$\! & $< 1.5 \cdot 10^{-8}$ \\
            \hline
            \!$\mathrm{BR}(\tau^- \!\to\! e^- \pi )$\! & $< 8.0 \cdot 10^{-8}$ \\
            \hline
            \!$\mathrm{BR}(\tau^- \!\to\! e^- \eta)$\! & $< 9.2 \cdot 10^{-8}$ \\
            \hline
            \!$\mathrm{BR}(\tau^- \!\to\! e^- \eta^\prime)$\! & $< 1.6 \cdot 10^{-7}$ \\
            \hline
        \end{tabular}
    \end{minipage}
    \hfill
    \begin{minipage}{0.48\linewidth}
        \centering
        \begin{tabular}{|c|c|}
            \hline
            \textbf{Observable} & \textbf{Constraint} \\
            \hline\hline
            \!$\mathrm{BR}(\tau^-$ $\to$ $\mu^- \pi )$\! & $< 1.1 \cdot 10^{-7}$ \\
            \hline
            \!$\mathrm{BR}(\tau^-$ $\to$ $\mu^- \eta)$\! & $< 6.5 \cdot 10^{-8}$ \\
            \hline
            \!$\mathrm{BR}(\tau^-$ $\to$ $\mu^- \eta^\prime)$\! & $< 1.3 \cdot 10^{-7}$ \\
            \hline
            \!$\mathrm{CR}_{\mu \to e}$(Ti)\! & $< 4.3 \cdot 10^{-12}$ \\
            \hline
            \!$\mathrm{CR}_{\mu \to e}$(Pb)\! & $< 4.3 \cdot 10^{-11}$ \\
            \hline
            \!$\mathrm{CR}_{\mu \to e}$(Au)\! & $< 7.0 \cdot 10^{-13}$ \\
            \hline
            \!$\mathrm{BR}(Z^0$ $\to$ $e^\pm \mu^\mp )$\! & $< 7.5 \cdot 10^{-7}$ \\
            \hline
            \!$\mathrm{BR}(Z^0$ $\to$ $e^\pm \tau^\mp )$\! & $< 5.0\cdot 10^{-6}$ \\
            \hline
            \!$\mathrm{BR}(Z^0$ $\to$ $\mu^\pm \tau^\mp)$\! & $< 6.5 \cdot 10^{-6}$ \\
            \hline
            \!$\mathrm{EDM}_e$\! & $< 1.1 \cdot 10^{-29} $
            \\
            \hline
            \!$\mathrm{EDM}_{\mu}$\! & $< 1.8 \cdot 10^{-19}$
            \\
            \hline
            \!$\mathrm{EDM}_{\tau}$\! & $< 1.15 \cdot 10^{-17} $
            \\
            \hline            
            \!$a_{e}^{\mathrm{BSM}}$~($\star$)\! & $\left(-87 \pm 103 \right) \cdot 10^{-14}$ \\
            \hline
            \!$a_{\mu}^{\mathrm{BSM}}$~($\star$)\! & $\left(251 \pm 59\right) \cdot 10^{-11}$ \\
            \hline
        \end{tabular}
        \vspace*{4.5mm}
    \end{minipage}
    \caption{Constraints from the Higgs-boson mass ($m_H$) \cite{Higgs2015}, the DM relic density ($\Omega_{\mathrm{CDM}}h^2$) \cite{Planck2018}, the anomalous magnetic moments of leptons $a_{\ell}^{\mathrm{BSM}}$ ($\ell=e,\mu$) \cite{ParticleDataGroup:2022pth}, the electric dipole moments of leptons EDM$_{\ell}$ ($\ell=e,\mu, \tau$) \cite{EDM_electron,EDM_muon,EDM_tau}, and numerous LFV and $CP$-violating observables (remaining entries) \cite{ParticleDataGroup:2022pth} as implemented in our MCMC and DNN analyses for the ``T1-2G'' and ``T1-2B ext.'' models. The first uncertainties are those used in the MCMC studies, while those in parentheses are the uncertainties used in the DNN studies (see text for details). The anomalous magnetic moments of leptons (marked by $\star$) are taken into account only in the studies of the ``T1-2G'' framework.}
    \label{Tab:Constraints}
\end{table}

As stated above, we will make use of the Markov Chain Monte Carlo (MCMC) technique, as well as the Deep Neural Network (DNN) technique, which are both well adapted to the exploration of a multi-dimensional parameter space in the presence of stringent constraints. We will present the two techniques in the following.

\subsection{Markov Chain Monte Carlo (MCMC)}
\label{Sec:Markov_Chain_Monte_Carlo}

The employed MCMC technique \cite{Markov1971} is an iterative scanning procedure, based on the Metropolis-Hastings algorithm \cite{Metropolis1953, Hastings1970}, where the parameter space exploration is conditioned -- in our case -- by a number of constraints. The latter are implemented through the computation of the likelihood associated to a given parameter set. 

Detailed explanations of the MCMC technique, and its application to scotogenic frameworks, can be found in Refs.\ \cite{Sarazin:2021nwo, Alvarez:2023dzz, deNoyers:2024qjz}. Consequently, we refer the reader to these references for a detailed discussion of the method and the associated likelihood computation and comparison. Running a number of MCMC chains allows to obtain posterior distributions for the scanned parameters, which can be compared, e.g., to the prior distributions, obtained from a pure random scan. Moreover, the posterior distributions allow to identify the phenomenologically viable parameter regions in view of the imposed constraints.

\subsection{Deep Neural Networks and Active Learning}
\label{Sec:DNN_ActiveLearning}

In the following, we will describe the technical details of the Deep Neural Networks (DNN) as well as the active learning strategy designed to efficiently select informative samples for labelling. This combination is particularly useful when labelled -- i.e.\ classified as ``good'' or ``bad'' parameter sets -- data is scarce or expensive to obtain.

Active learning is an iterative approach aimed at reducing labelling costs by selectively querying the most informative data points from an unlabelled dataset. The active learning cycle typically proceeds as follows:
\begin{itemize}
    \item \textbf{Initial training:} a DNN is first trained on a small labelled dataset $L$;
    \item \textbf{Query strategy:} an informativeness measure is used to select the most informative samples from an unlabelled dataset $U$. These samples will have a positive impact in the improvement of the model if they are labelled;
    \item \textbf{Oracle query:} the labels for the selected data points are requested from an oracle;
    \item \textbf{Model update:} the labelled dataset $L$ is augmented with the newly labelled samples, and the DNN is retrained or fine-tuned;
    \item \textbf{Iteration:} the process is repeated until the desired performance is reached or the labelling budget is exhausted.
\end{itemize}
In our scotogenic case study, this general procedure is implemented based on the discussion in Ref.\ \cite{Goodsell2022}.

The first step is to choose the exact architecture of the DNN to be used in the study. To this end, we have performed hypertuning over the number of layers, the number of neurons per layer, the activation function, and the learning rate. We have used the dataset of the MCMC and random analyses of the ``T1-2G'' model to perform this hypertuning. The architecture that turned out to be most efficient is illustrated in Fig.\ \ref{Fig:DNN_Architecture}.

\begin{figure}
    \centering
    \begin{tikzpicture}[scale=1.0]
        \def\h{-1.2} 
        \def\w{2.8} 
  
        \node[mybox=myred] (0A) at (-3*\w,0) {%
            21 Input};

        \node[mybox=myblue] (0B) at (-2*\w,0) {%
            320 nodes};

        \node[mybox=myblue] (0C) at (-\w,0) {%
            128 nodes};

        \node[mybox=myblue] (1C) at (0,0) {%
            224 nodes};

        \node[mybox=myblue] (1B) at (\w,0) {%
            96 nodes};

        \node[mybox=myblue] (1A) at (\w,\h) {%
            512 nodes};

        \node[mybox=myblue] (2A) at (0,\h) {%
            224 nodes};

        \node[mybox=myblue] (2B) at (-\w,\h) {%
            384 nodes};

        \node[mybox=myblue] (2C) at (-2*\w,\h) {%
            32 nodes};

        \node[mybox=myblue] (3C) at (-3*\w,\h) {%
            192 nodes};

        \node[mybox=myblue] (3B) at (-3*\w,2*\h) {%
            128 nodes};

        \node[mybox=myblue] (3A) at (-2*\w,2*\h) {%
            64 nodes};

        \node[mybox=myblue] (4A) at (-\w,2*\h) {%
            64 nodes};

        \node[mybox=mygreen] (4B) at (0,2*\h) {%
            1 Output};

        \draw[myarrow=myred] (0A) -- (0B);
        \draw[myarrow=myblue]  (0B) -- (0C);
        \draw[myarrow=myblue]  (0C) -- (1C);
        \draw[myarrow=myblue]  (1C) -- (1B);
        \draw[myarrow=myblue]  (1B) -- (1A);
        \draw[myarrow=myblue]  (1A) -- (2A);
        \draw[myarrow=myblue]  (2A) -- (2B);
        \draw[myarrow=myblue]  (2B) -- (2C);
        \draw[myarrow=myblue]  (2C) -- (3C);
        \draw[myarrow=myblue]  (3C) -- (3B);
        \draw[myarrow=myblue]  (3B) -- (3A);
        \draw[myarrow=myblue]  (3A) -- (4A);
        \draw[myarrow=mygreen] (4A) -- (4B);
  
    \end{tikzpicture}
    \caption{Schematic representation of our DNN architecture. Each layer is a dense layer, fully connected and feed forward.}
    \label{Fig:DNN_Architecture}
\end{figure}
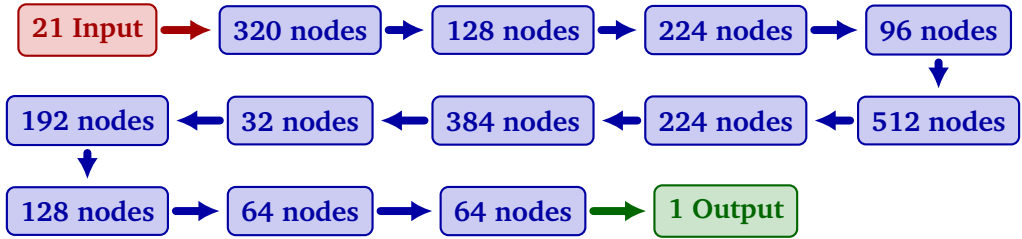

\begin{figure}
    \resizebox{\columnwidth}{!}{%
    \begin{tikzpicture}[align=center]
    \xdefinecolor{Atlas_green}     {HTML}{4cb944}
    \xdefinecolor{Atlas_red}       {HTML}{f4442e}
    \xdefinecolor{Atlas_yellow}    {HTML}{ffb800}
    \xdefinecolor{Atlas}{HTML}{0b80c3}

    \node[trapezium,
        draw,
        thick,
        trapezium stretches = true,
        minimum width = 3cm, 
        minimum height = 1.5cm,
        rotate=270] (generateur) at (0,0) {\rotatebox{90}{Discriminator}};
    
    \draw[Atlas_red, fill, very thick] (1.26,-0.23) rectangle (1.27,0.23);
    \draw[red] (1.265,-0.5) node{$\sigma$};
    
    \node (logit) at (2.4,0) {$\hat{x}^i \in [0;1]$};
    \node (norm) at (-3,4) {$x = \left(x_{\rm max} - x_{\rm min}\right) \tilde{x} + x_{\rm max}$};
    \draw [-latex] (norm.east) to (0.75,4);
    
    \node [draw, 
        Atlas,
        thick,
        minimum width=1.2cm,
        minimum height=0.5cm,
    ]  (oracle) at (1.5,4) {Oracle}; 

    \node (loss) at (1,2.5) {$\mathcal{C}_{\rm BCE}$};
    
    \draw [-latex] (oracle.south) to (loss.north);
    \draw [-latex] (logit.north) to (loss.south);
    \draw [-latex] (loss.west) to [in=30,out=180] (generateur.west);
    
    \node (q) at (6,2.5) {$q$ = \small{wrong classification (\%)}};
    
    \draw [-latex] (logit.north) to [in=180] (q.west); 
    \draw [-latex] (oracle.east) to [out=0,in=90] (q.north); 
    
    \node (p) [below=of q] {$p= 2 \min(q,0.5)$};
    \draw [-latex] (q.south) to (p.north); 
    
    \node (pr) [below= 3 of p] {$pK$ \textit{random} \\ $(1-p)K$ near good points};
    \draw [-latex] (p.south) to (pr.north); 

    \node (input) at (-3.75,0) {
    $\begin{bmatrix}
        \tilde{x}_1^1 \\ . \\ . \\ . \\ . \\ \tilde{x}_{N}^1
    \end{bmatrix}$ 
    $\begin{matrix}
    & . & . & . &  \\ \\ \\ & . & . & . & \\ \\ \\ & . & . & . &
    \end{matrix}$
    $\begin{bmatrix}
        \tilde{x}_1^M \\ . \\ . \\ . \\ . \\ \tilde{x}_{N}^M
    \end{bmatrix}$ 
    };
    
    \draw [-latex] (input.north) to [out=90,in=270] (norm.south);

    \draw [latex-latex,Atlas_green] (-2,-2) to (-3.70,-2);
    \draw [Atlas_green] (-2.2,-2.5) node{$pK$ times fully \\ picked from $\mathcal{U}_{[0;1]}$};
    \draw [-latex,Atlas_green] (4.75,-2.75) to (-1,-2.5); 
    
    \draw [latex-latex, Atlas_yellow] (-3.70,-2) to (-5.4,-2);
    \draw [Atlas_yellow] (-4.5,-2.5) node{$(1-p)K$};
    
    \node (L) at (3,-4.5) {generate $L \gg K$ \\ fully picked from $\mathcal{U}_{[0;1]}$};
    \draw [-latex,Atlas_yellow] (3.75,-3.25) to [in=90,out=180] (L.north);
    
    \node (L2) at (-3,-4) {with a scoring system \\ we select points near the good points};
    \draw [-latex,Atlas_yellow] (L.west) to [in=0,out=180] (L2.east);
    \draw [-latex,Atlas_yellow] (L2.north) to [in=270,out=90] (-4.5,-3);
    
    \end{tikzpicture}
    }
    \caption{Schema showing the Active Learning process. The input data, of size $K$, are both given to the Discriminator (DNN) and the Oracle (\texttt{SPheno} and \texttt{MicrOMEGAs}). Based on the output we take $pK$ fully randomly and $(1-p)K$ near precedent good points based on a scoring methods.}
    \label{Fig:SchemeAL}
\end{figure}
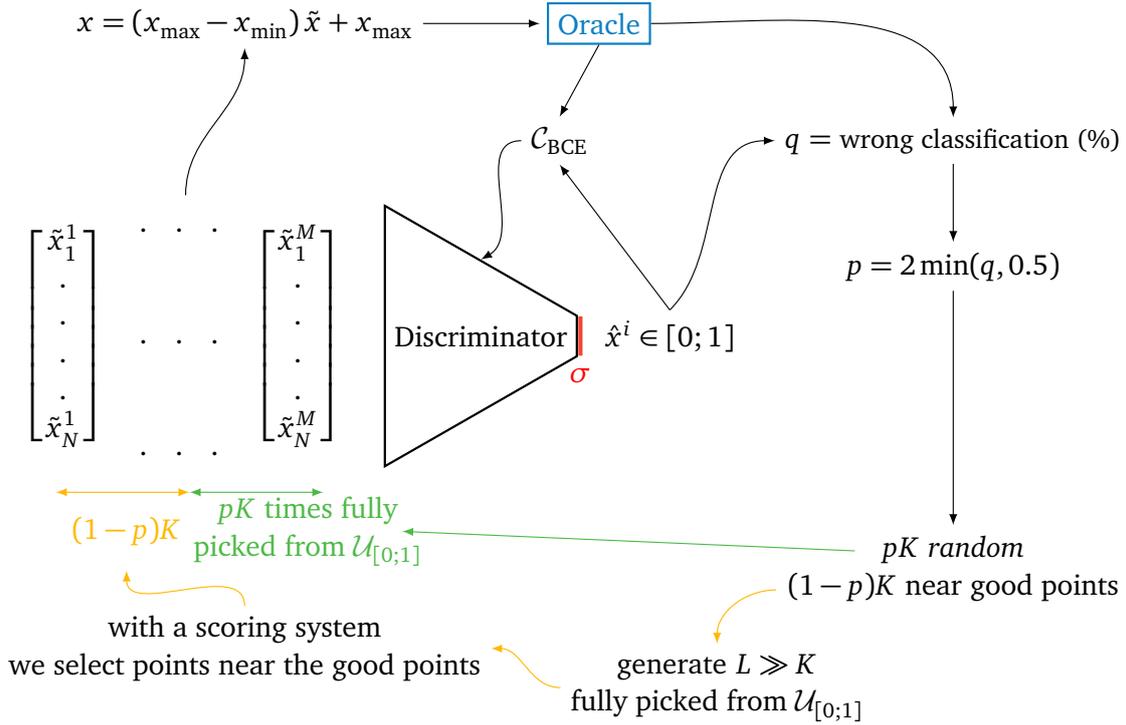

The chosen DNN is composed of 13 fully connected layers, with the first layer having 21 neurons and the last layer having only one neuron. As activation function, we have chosen to use the rectifier linear unit (ReLU) function for each layer apart from the last layer which has a sigmoid activation function. As optimizer, we have chosen the NADAM optimizer \cite{dozat.2016} provided by the deep learning framework \texttt{Keras} \cite{chollet2015keras} from the \texttt{Python} library \texttt{TensorFlow} \cite{tensorflow2015-whitepaper}. The learning rate, important for the update of weights and biases of the DNN via the backpropagation process, is set to $\alpha = 0.01$. The ReLU and sigmoid $\sigma$ functions are defined as
\begin{equation}
    \mathrm{ReLU}(x) ~=~ \max(0, x) \,, \quad{\rm and}\quad \sigma(x) ~=~ \frac{1}{1 + e^{-x}} \,.
    \label{Eq:DNN:Activation_functions}
\end{equation}

The loss function used is the binary cross-entropy loss $\mathcal{C}_{\mathrm{BCE}}$, defined as
\begin{equation}
    \mathcal{C}_{\mathrm{BCE}} = -\frac{1}{N} \sum_{i=1}^{N} \Bigl[y_i \log(\hat{y}_i) + (1-y_i) \log(1-\hat{y}_i)\Bigr],
    \label{Eq:DNN_Loss_function}
\end{equation}
where $y$ is the expected value and $\hat{y}$ is the value predicted by the DNN. In a binary problem, the expected value $y$ is either $0$ or $1$, $0$ being the wrong case and $1$ the correct one. In case of $y = 0$ the loss will be quantified by the second term of Eq.~\eqref{Eq:DNN_Loss_function}, and in case of $y = 1$ the loss will be quantified by the first term of Eq.~\eqref{Eq:DNN_Loss_function}. The loss function is then averaged over the number of samples $N$. This allows us to get an estimate of the quality of the DNN. If the obtained value is too high, this means that the DNN makes too many mistakes classifying the points. Once the loss value is computed, the DNN will update its weights and biases. During the use of the DNN, there are three phases. The first phase is the training phase, in which the DNN is trained on a labelled dataset. The second phase is the testing phase, in which we will give to the DNN a set of points to classify and then optimizing its variables in order to minimize its loss. The third phase is the using phase, during which we will use the trained DNN to classify new data.

\subsubsection*{Active Learning strategy}
\label{Sec:AL_strategy}

This part of the discussion is based on Ref.\ \cite{Goodsell2022} and uses Active Learning (AL). The DNN will use the output of an Oracle, that will classify the data to train the DNN. The purpose here is to train a parameter set discriminator. In both the ``T1-2G'' and the ``T1-2B ext.'' models, we have $21$ parameters to discriminate. For the following of this Section, $\mathcal{P}$ will refer to the number of discriminated parameters in the ``T1-2G'' and ``T1-2B ext.'' model. These parameters are listed in Table \ref{Tab:ScanParameters} with their respective interval. The Active Learning process is depicted schematically in Fig.\ \ref{Fig:SchemeAL}

\subsubsection*{Oracle scoring system}
\label{Sec:Oracle_scoring_system}

\begin{figure}[tb]
\centering
\includegraphics[width=0.9\textwidth]{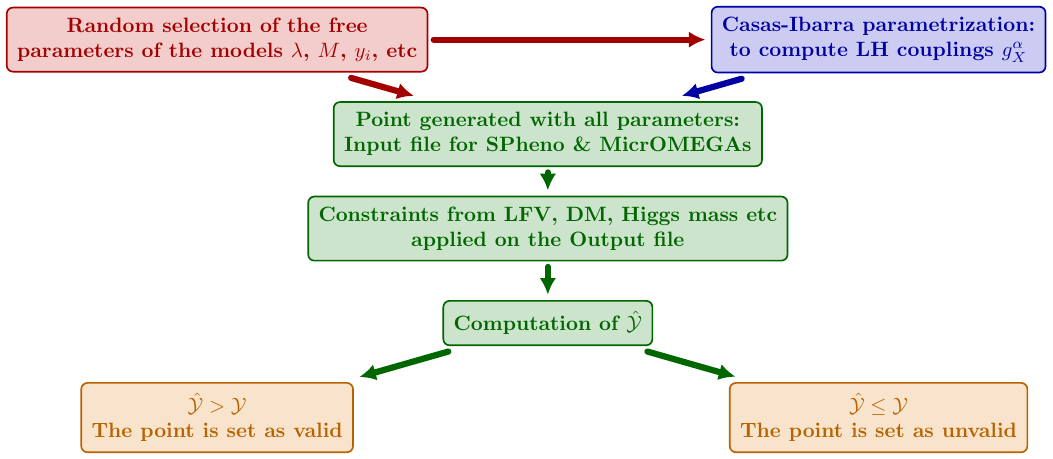}
\caption{A simple flowchart illustrating the AL pipeline for ``T1-2G'' model and ``T1-2B ext.'' model.}
\label{Fig:AL_pipeline}
\end{figure}

In order to determine the ``quality'' of a parameter set, one gives the latter to the Oracle, which will classify the parameter set as either good or bad based on a condition that we choose. The procedure is depicted schematically in Fig.\ \ref{Fig:AL_pipeline}. In the Oracle we use \texttt{SPheno} \cite{SPheno2003, SPheno2011} and \texttt{MicrOMEGAs} \cite{MO2001, MO2004, MO2007a, MO2007b, MO2013, MO2018} to compute the physical spectrum and all the observables listed in Table \ref{Tab:Constraints}. Then we compute the ordinate $\mathcal{\hat{Y}}_i$ for each considered observable $\mathcal{\hat{X}}_i$ according to their prior distribution, e.g.\ if the Higgs mass given be \texttt{SPheno} is equal to the mean value given in Table \ref{Tab:Constraints} we will set $\mathcal{\hat{Y}}_h = 1$. The value of $\mathcal{\hat{Y}}_h$ will get closer to $0$ as the Higgs mass given by \texttt{SPheno} gets further from the mean value. The same procedure is applied to all the observables listed in Table \ref{Tab:Constraints}. The final score $\mathcal{\hat{Y}}$ is then computed as
\begin{equation}
    \mathcal{\hat{Y}} = \left( \prod_{i=1}^{N} \mathcal{\hat{Y}}_i \right)^{\!\!\frac{1}{N}}.
        \label{Eq:CompTools:Score_function_after_SPheno_MO}
\end{equation}
The choice of the geometric mean is made in order to avoid the score to ignore completely one of the observables, considered that only one will be off the charts. 

We do that same procedure for the ordinate $\mathcal{Y}_i$ associated with the upper bounds of the observables. Going back to the example of the Higgs mass, the ordinate $\mathcal{Y}_h$ will be associated to the abscissa $\mathcal{X}_h$ of the upper bound of the Higgs mass, i.e.\ $m_h = 140 \, \mathrm{GeV}$ in our case. For the quantities that are not Gaussian distributed but rather step functions, we will allow $50\%$ tolerance and we compute that $\mathcal{Y}_i$ at this $50\%$ tolerance threshold. The score $\mathcal{Y}$ is then computed as
\begin{equation}
    \mathcal{Y} ~=~ \left( \prod_{i=1}^{N} \mathcal{Y}_i \right)^{\!\!\frac{1}{N}}.
        \label{Eq:CompTools:Score_function_upper_bounds}
\end{equation} 

To determine whether the point is classified as ``good'' or ``bad'' by the Oracle, we compute the global score $\mathcal{S}$ as
\begin{equation}
    \mathcal{S} ~=~ \frac{\mathcal{\hat{Y}}}{\mathcal{Y}} \,.
    \label{Eq:CompTools:Score_function_ratio}
\end{equation}
This specific choice of scoring function in the Oracle is made in order to be able to compare the score of the points with the score of the upper bounds. The Oracle will classify the point as ``good'' if $\mathcal{S} > 1$ and as ``bad'' if $\mathcal{S} \leq 1$.

\subsubsection*{DNN scoring system}
\label{Sec:DNN_scoring_system}

Going back to the discussion about the discriminator, we train our DNN over $N$ steps. For each training step we give it $K \cdot \mathcal{P}$ normalized parameters. As feedback we get $K$ predictions from the DNN, $K \in \left[0, 1 \right]$. One can notice here that we are not in a binary problem, leading to $K$ being a value between $0$ and $1$ instead of being strictly $0$ or $1$. First we define a quantity $p = 2\min(q, 0.5)$, where $q$ is the percentage of wrong classification. In case of $p=1$ our DNN does not work better than a random classifier, then we choose our inputs $K \in \left[0,1\right]$ according to an uniform distribution $\mathcal{U}_{\left[0,1\right]}$. In case of $p < 1$, we choose $pK$ inputs randomly from the $\mathcal{U}_{\left[0,1\right]}$ distribution and $(1-p)K$ inputs near the good points in order to explore the possible physical boundaries of the parameter space. To do so, we have a set of $L \cdot \mathcal{P}$ parameters, chosen also according to the uniform distribution $\mathcal{U}_{\left[0,1\right]}$. More details about the scoring system can be found in Ref.\ \cite{Goodsell2022}. 
\section{Results and discussion}
\label{Sec:Results}
In this Section, we are going to discuss our results on the level of the phenomenology of the considered scotogenic models. A more technical comparison of the two employed scanning techniques will follow in Section \ref{Sec:Comparison}. In this first comparative study, we limit ourselves to the case of fermionic dark matter, i.e.\ the lightest fermion $\chi^0_1$ being the dark matter candidate within both the ``T1-2B ext.'' and ``T1-2G'' model. As already shown in Refs.\ \cite{Sarazin:2021nwo, deNoyers:2024qjz}, fermionic dark matter has an overall higher likelihood as compared to scalar dark matter within comparable scotogenic frameworks under consideration. A discussion of phenomenological aspects of scalar dark matter within the ``T1-2G'' model can be found in Ref.\ \cite{deNoyers:2024qjz}.

\subsection{The ``T1-2G'' framework}
\label{Sec:ResultsT12G}

The constraints stemming from the neutrino sector being satisfied through the application of the Casas-Ibarra parametrization discussed in Sec.\ \ref{Sec:NeutrinoMassGeneration}, the most important constraint within the scotogenic frameworks under consideration is the dark matter relic density. As already discussed in Ref.\ \cite{deNoyers:2024qjz}, the relic density constraint can be met for specific parameter configurations, especially where co-annihilations dominate the total dark matter annihilation cross-section. In this configuration, the key parameter governing the co-annihilation cross-section, and thus the relic density, is the mass of the dark matter candidate. 

\begin{figure}
    \centering
    \includegraphics[width=0.76\textwidth]{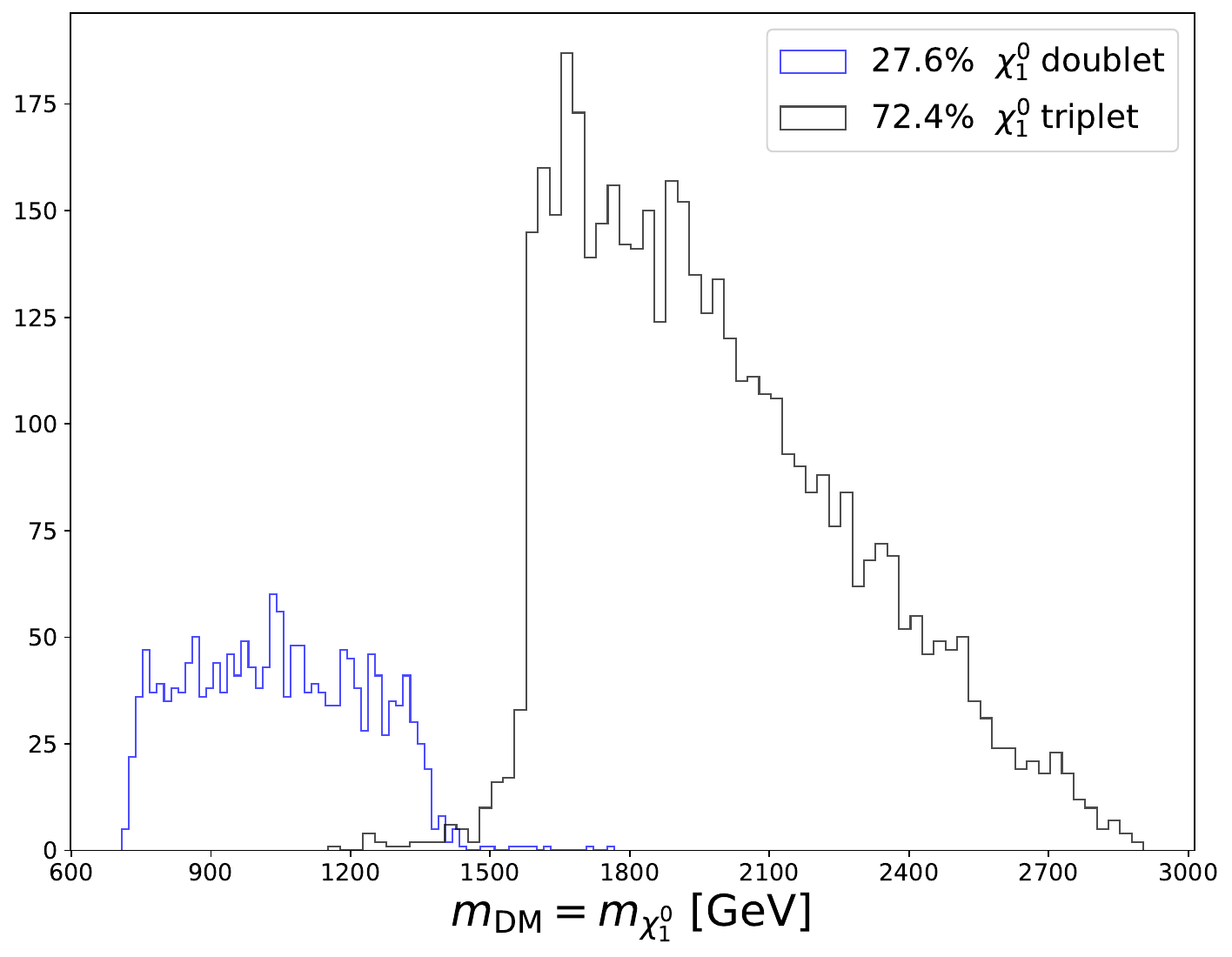}
    \caption{Histograms of the DM masses obtained from the DNN analysis, within the ``T1-2G'' model. The colour code indicates the DM nature in terms of fermionic doublet or triplet domination.}
    \label{Fig:Histo_DM_T12G}
\end{figure}

In Fig.\ \ref{Fig:Histo_DM_T12G}, we display the mass distribution obtained from our Deep Neural Network (DNN) analysis, imposing the relic density constraint of Table \ref{Tab:Constraints}. Two distinct viable parts are observed, the first spanning the interval between about 800 GeV and 1400 GeV corresponding to doublet-like dark matter, the second spanning the interval between about 1500 GeV and about 2900 GeV, corresponding to triplet-like dark matter. In both cases, the relic density constraint is met through important co-annihilations with the other components of the doublet or triplet, respectively. 

Comparing with the results presented in Ref.\ \cite{deNoyers:2024qjz}, it can be seen that the DNN analysis leads to similar conclusions as the Markov Chain Monte Carlo (MCMC) method when applied to the same model and constraints. The distribution shapes are somewhat different than the rather distinct peaks obtained from the MCMC study can be traced to the fact that the present study features less valid parameter points (about 6300 compared to 117\,000 for the MCMC study of Ref.\ \cite{deNoyers:2024qjz}) and the allowed interval for the dark matter relic density is chosen to be much less restrictive in the present study (see Table \ref{Tab:Constraints}). Moreover, the AL algorithm is not designed to converge to local optima but rather to map out the boundaries of the parameter space of the model.

Note also that the percentages given in the legend of Fig.\ \ref{Fig:Histo_DM_T12G} cannot directly be compared to the numbers given in Ref.\ \cite{deNoyers:2024qjz}, as the present study does not cover the case of scalar dark matter. The less good resolution of the distributions in Fig.\ \ref{Fig:Histo_DM_T12G} is explained by the lower number of points in the present study as compared to the MCMC study of Ref.\ \cite{deNoyers:2024qjz}. In addition, let us recall that the imposed intervals for the different constraints are less restrictive than in the MCMC study of Ref.\ \cite{deNoyers:2024qjz}, and the present DNN includes a larger random component.

\begin{figure}
    \centering
    \includegraphics[width=0.8\textwidth]{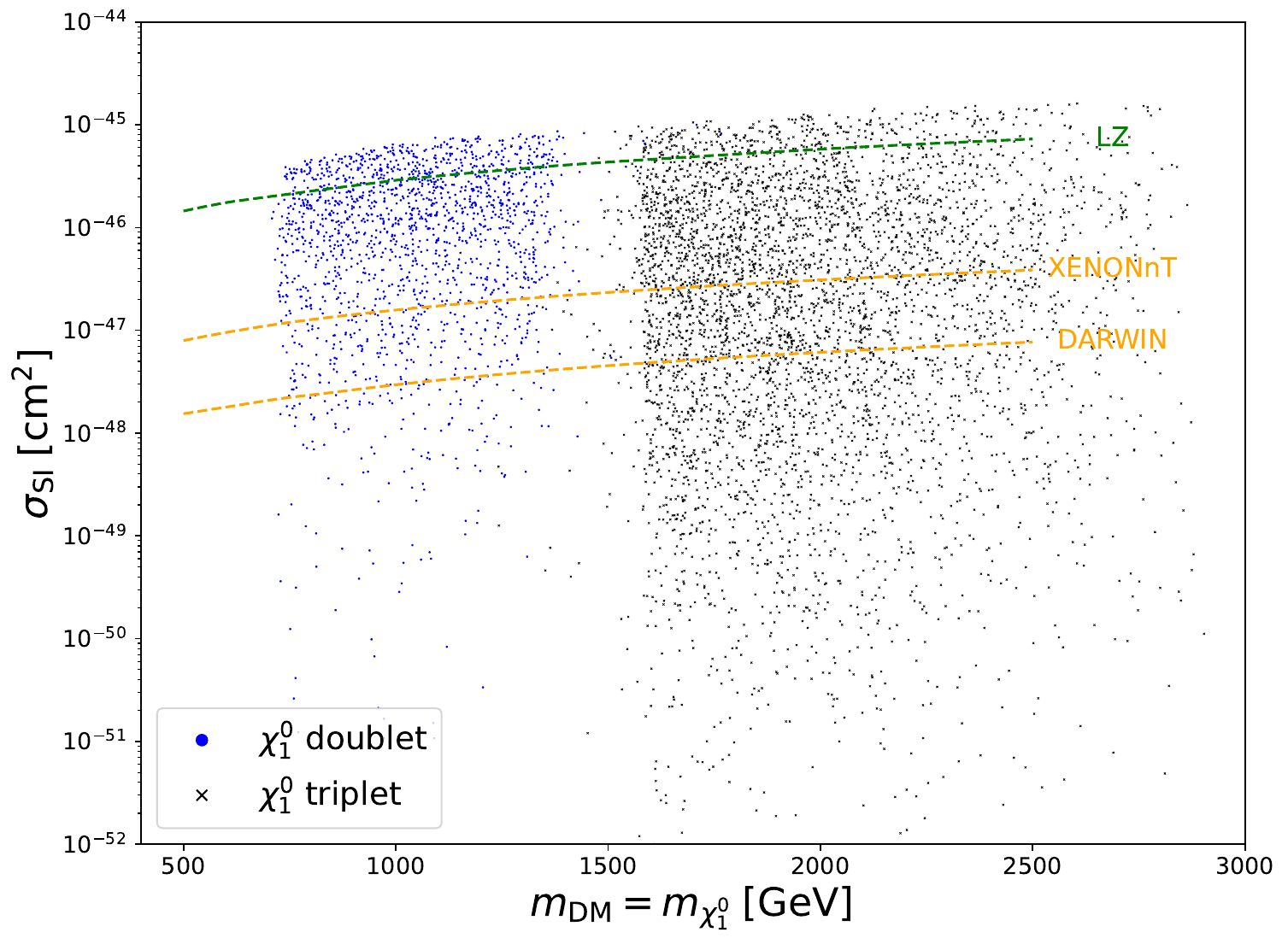}
    \caption{Spin-independent dark matter direct detection cross-section as a function of the dark matter mass obtained from the DNN analysis, within the ``T1-2G'' model.}
    \label{Fig:DMDD_T12G}
\end{figure}

Fig.\ \ref{Fig:DMDD_T12G} shows the spin-independent dark matter direct detection cross-section as a function of the dark matter mass for the phenomenologically viable parameter sets obtained from the DNN analysis. As can be seen, a large majority of the viable points are relatively close to the imposed LZ limit, and may consequently be challenged by the upcoming dark matter searches such as XENONnT and DARWIN. Note that single points may feature a cross-section slightly above the imposed limit, which is -- in a similar way as for an MCMC analysis of Ref.\ \cite{deNoyers:2024qjz} -- an artefact of the employed scanning technique.

\begin{figure}
    \centering
    \includegraphics[width=0.49\textwidth]{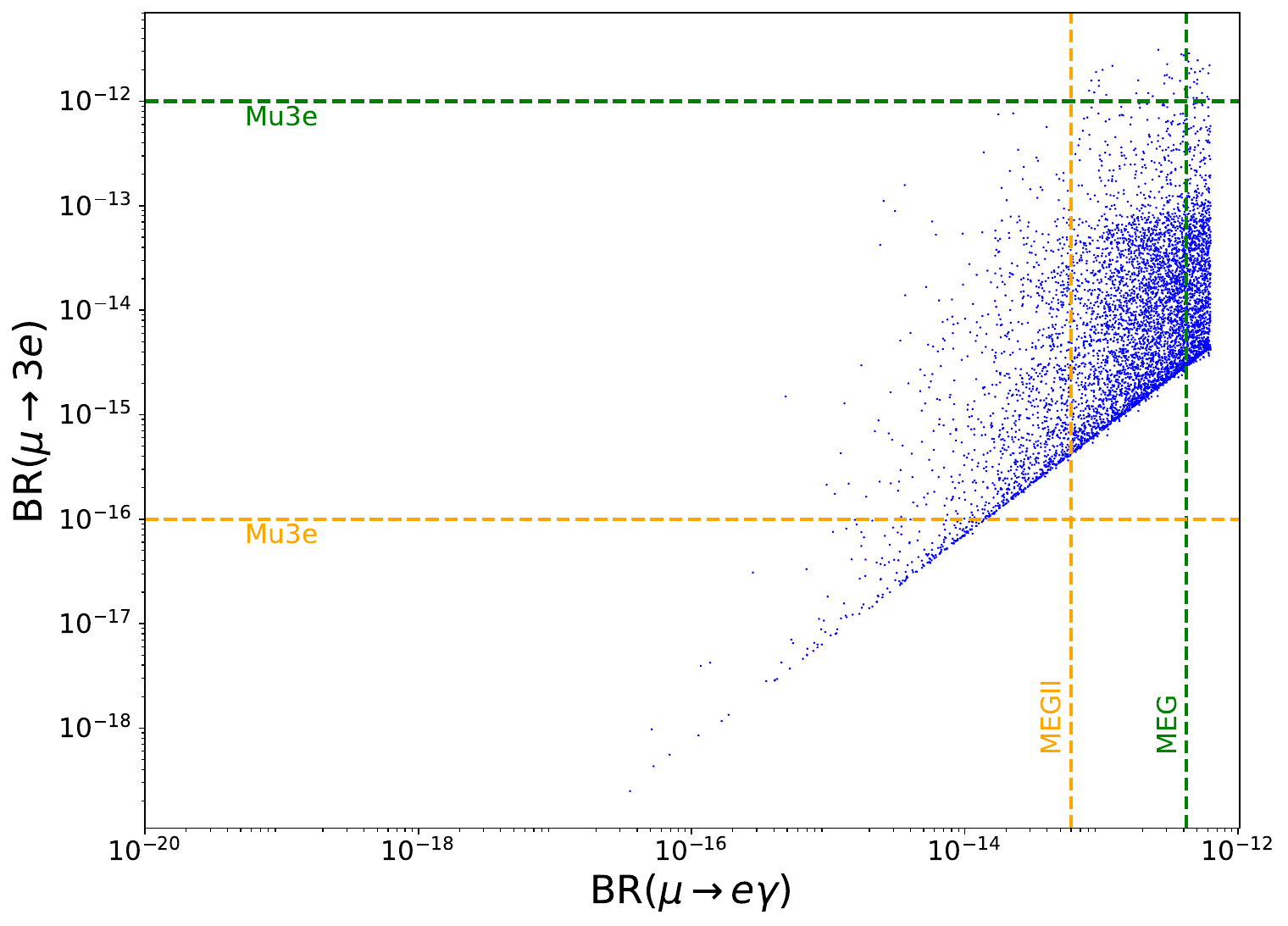}
    \includegraphics[width=0.49\textwidth]{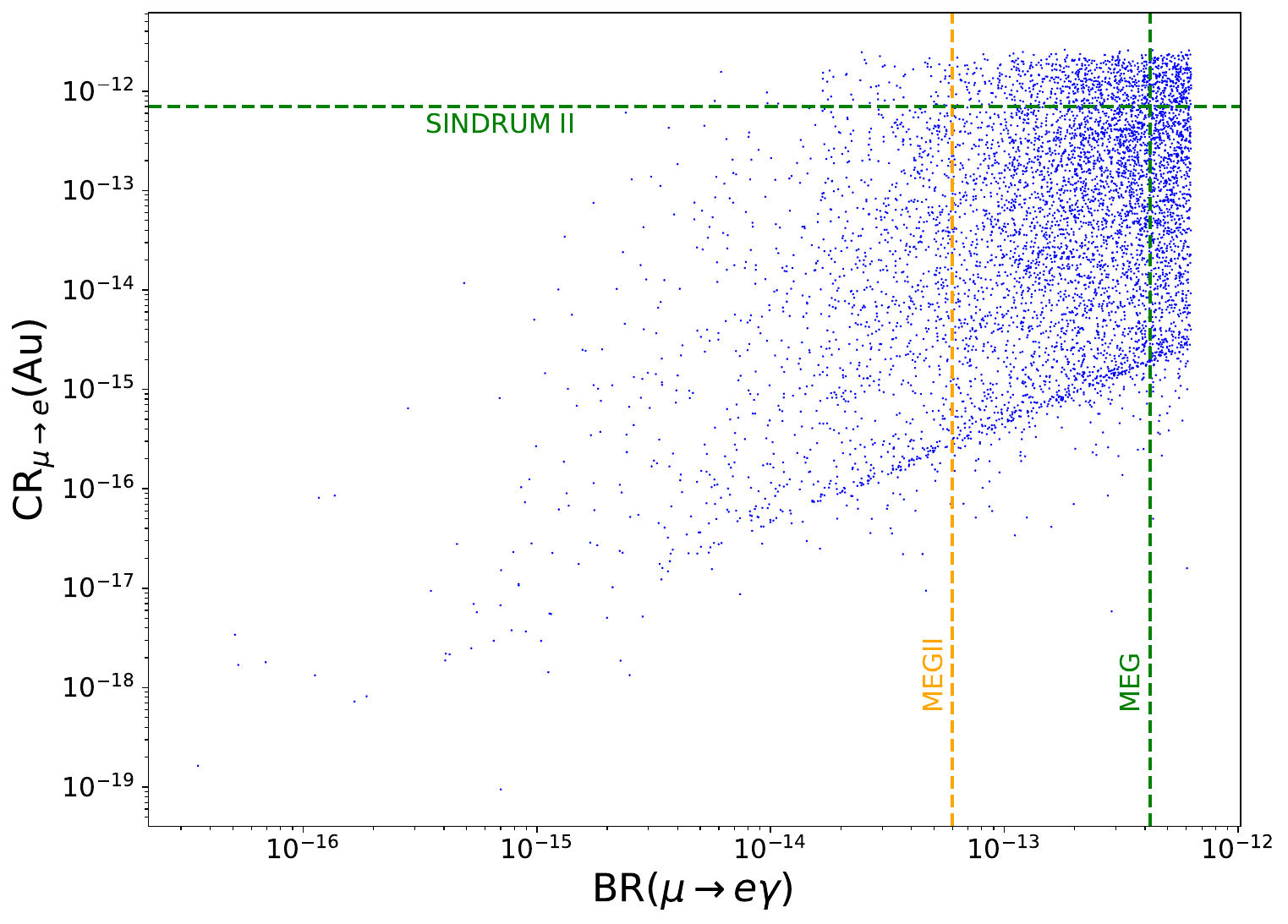}
    \caption{Correlations between the branching ratio of $\mu\to e\gamma$ and the branching ratio of $\mu\to 3 e$ (left) and the $\mu-e$ conversion rate within gold atoms (right) obtained from the DNN analysis in the ``T1-2G'' scotogenic framework.}
    \label{Fig:LFV_T12G}
\end{figure}

Finally, in Fig.\ \ref{Fig:LFV_T12G}, we display the branching ratios of the rare leptonic decays $\mu\to e\gamma$ and $\mu\to 3e$ as well as the $\mu-e$ conversion rate in gold as three examples of lepton flavour changing observables. Again, few points are slightly over the imposed limits due to the scanning technique. We observe the usual \cite{Sarazin:2021nwo, Alvarez:2023dzz, deNoyers:2024qjz} correlations between the different observables related to $\mu-e$ transitions, which become washed out at numerically larger coupling values. Overall, as can be seen, a large part of the currently viable parameter space will be challenged by future precision measurements, especially the MEG II and COMET\footnote{COMET: search for conversion in Al, not shown in Fig.\ \ref{Fig:LFV_T12G}.} experiments.

\subsection{The ``T1-2B ext.'' framework}
\label{Sec:ResultsT12Bext}

In this section, the results obtained from the AL scan for the ``T1-2B ext.'' model will be presented. A total of 2603 viable points featuring fermionic dark matter were obtained.

\begin{figure}
    \centering
    \includegraphics[width=0.76\textwidth]{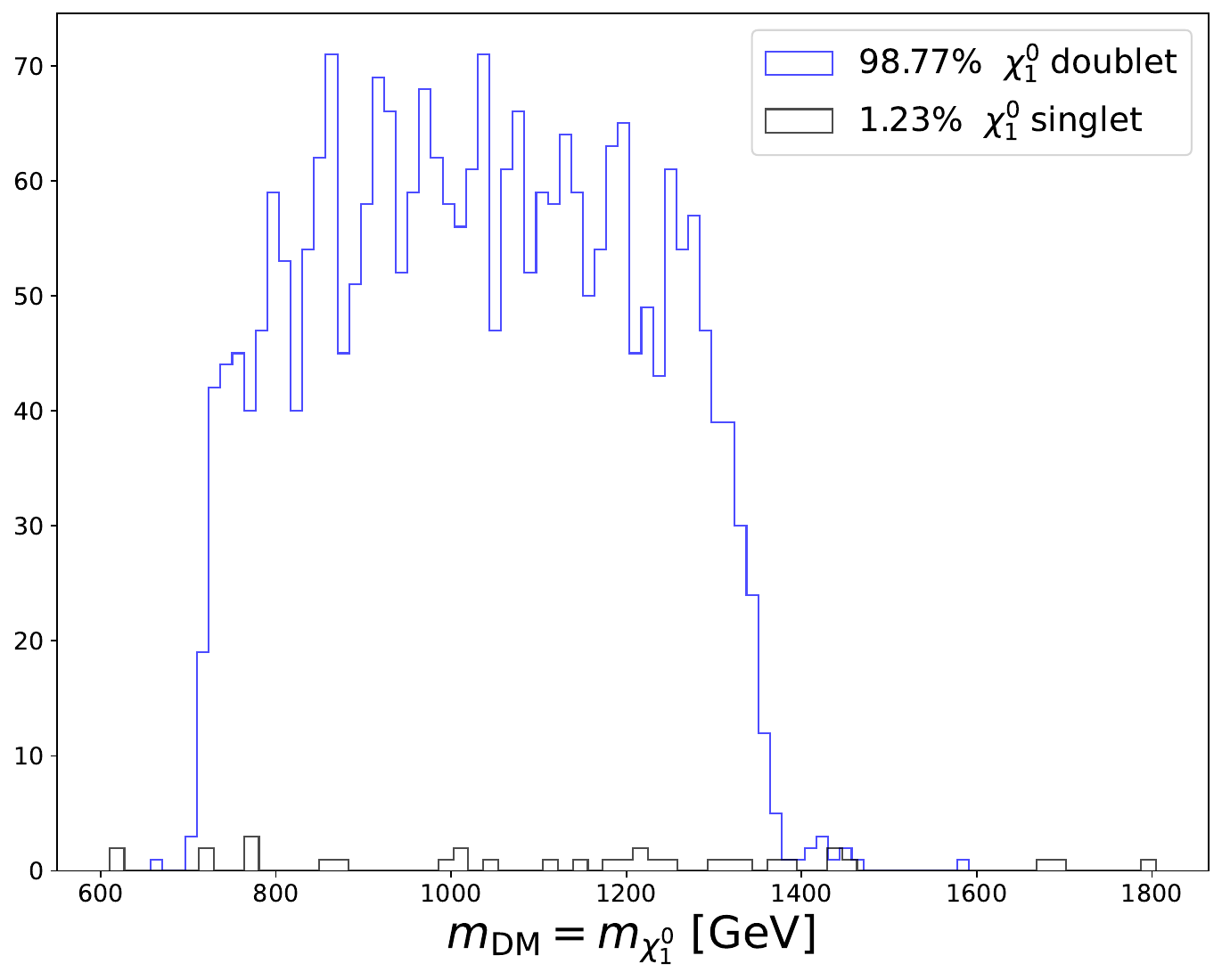}
    \caption{Histograms of the DM masses obtained from the DNN analysis, within the ``T1-2B ext.'' model. The colour code indicates the DM nature in terms of fermionic singlet or doublet domination.}
    \label{Fig:Histo_DM_T12Bext}
\end{figure}

As for the ``T1-2G'' framework, we start the discussion of the ``T1-2B ext.'' framework with the distribution of the dark matter mass depicted in Fig.\ \ref{Fig:Histo_DM_T12Bext}, as obtained from the Deep Neural Network (DNN) approach including the relic density constraint. In this case, only one peak is observed, corresponding to the situation where the dark matter particle is doublet-like and achieves the relic density through co-annihilations within the doublet. Note that singlet-like dark matter is highly unlikely, the only cases being configurations where co-annihilations with either the fermionic doublet or the scalars are possible. 

Coming to dark matter direct detection, the spin-independent cross-section is found, for a majority of valid parameter points, between the current LZ limit and the projected XENONnT and DARWIN reaches, meaning that a large part of the parameter space will be challenged by these experiments in a very near future.

\begin{figure}
    \centering
    \includegraphics[width=0.8\textwidth]{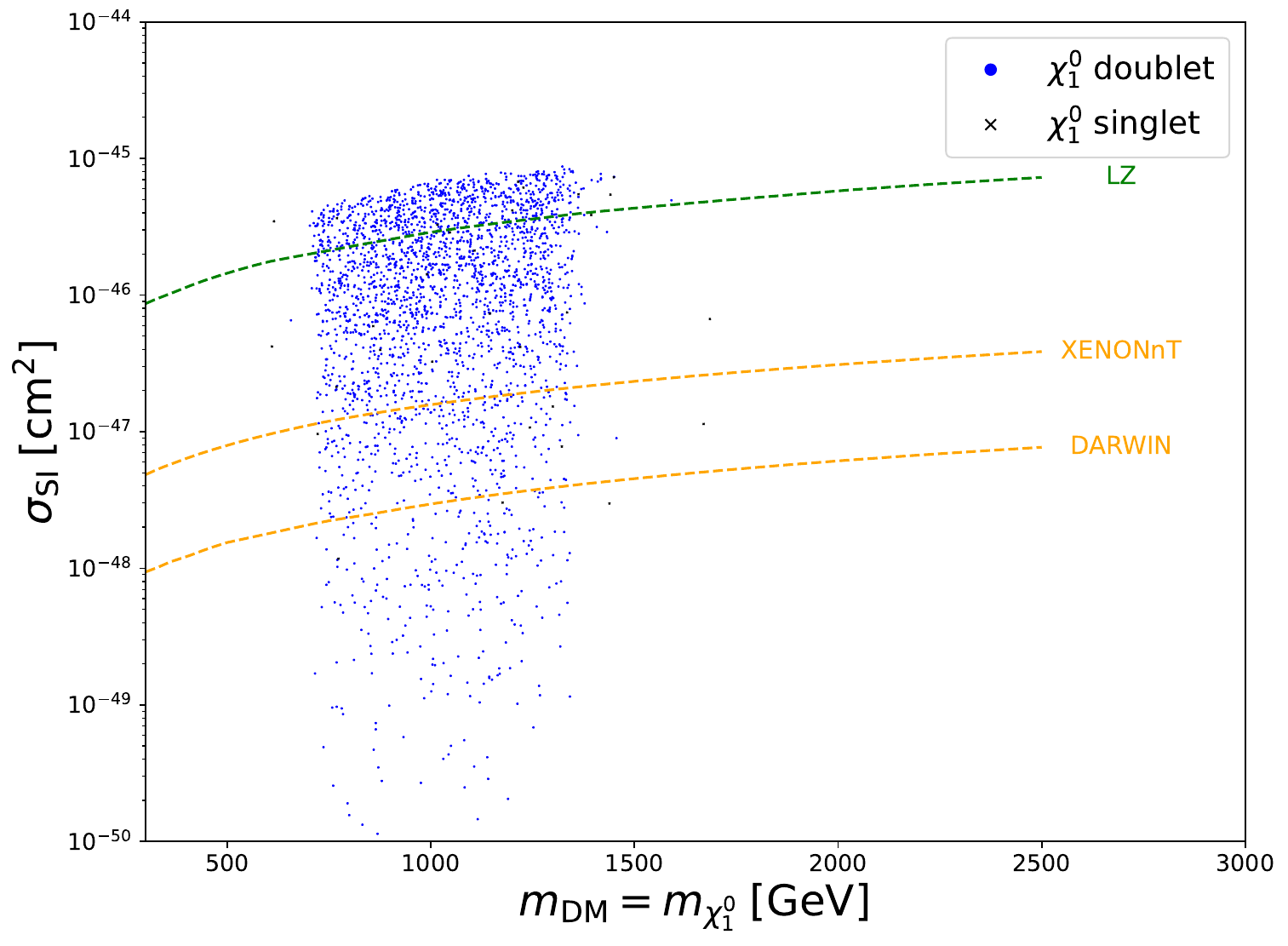}
    \caption{Spin-independent dark matter direct detection cross-section as a function of the dark matter mass obtained from the DNN analysis, within the ``T1-2B ext.'' model.}
    \label{Fig:DMDD_T12Bext}
\end{figure}

\begin{figure}
    \centering
    \includegraphics[width=0.49\textwidth]{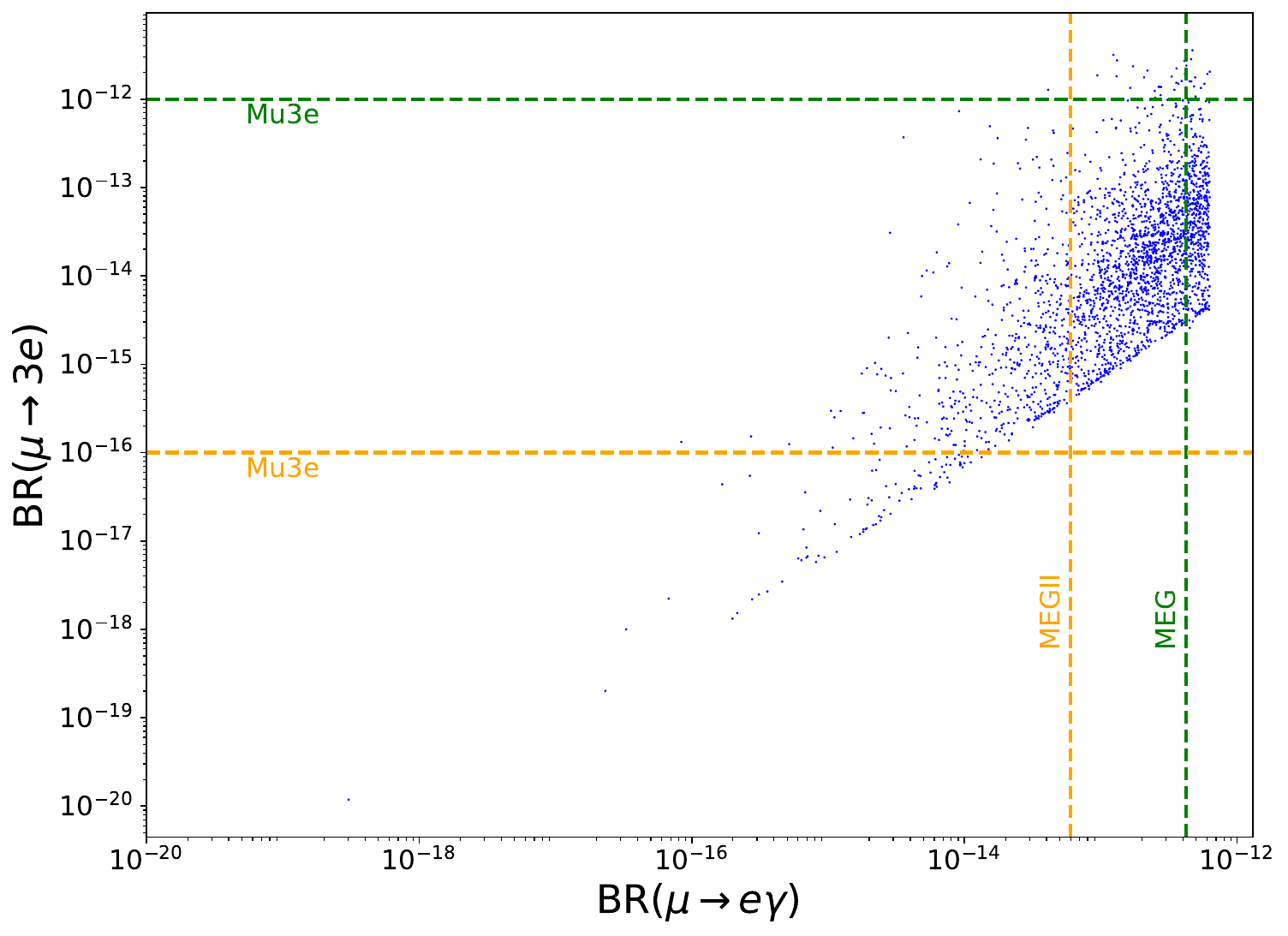}
    \includegraphics[width=0.49\textwidth]{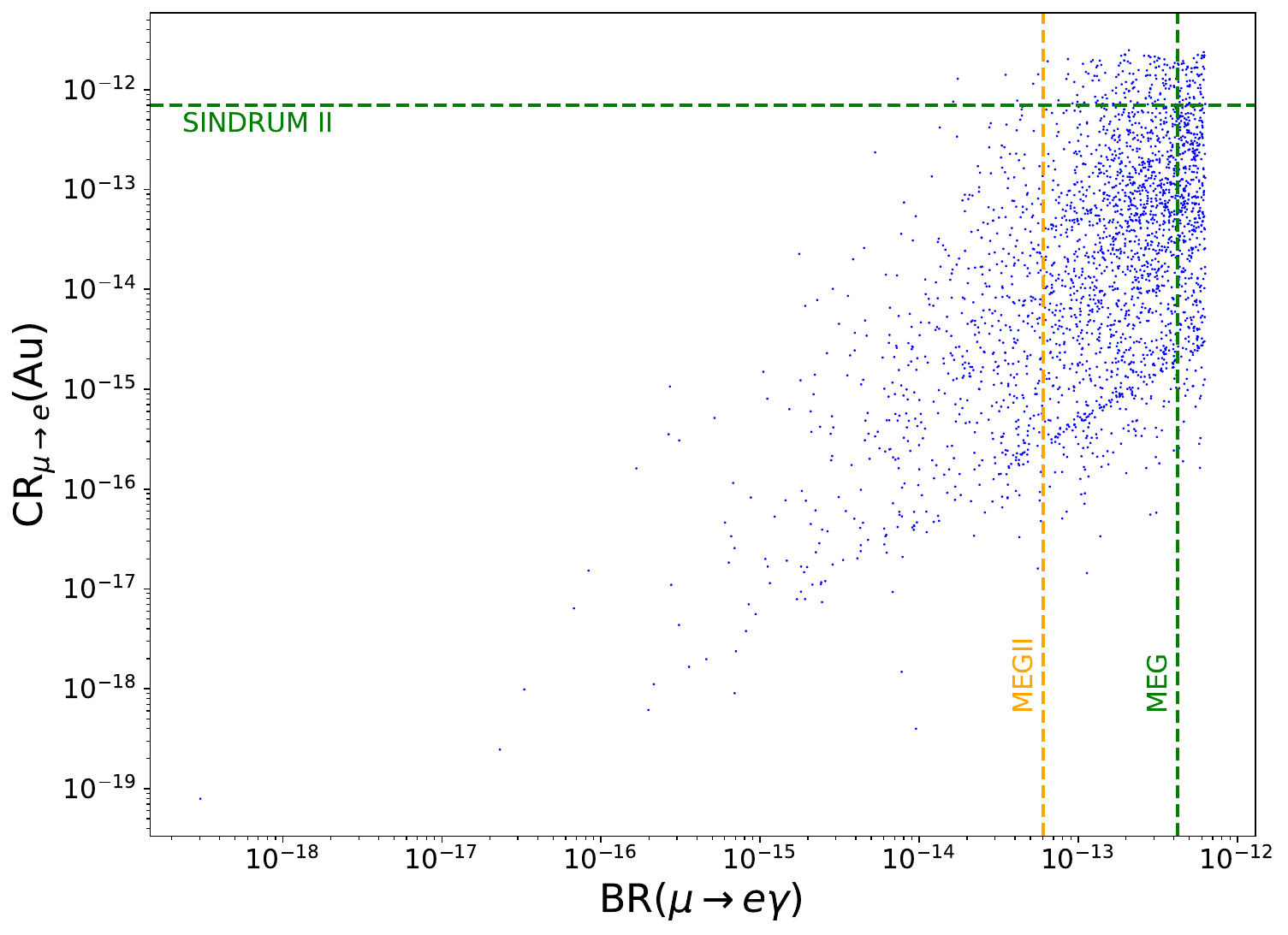}
    \caption{Correlations between the branching ratio of $\mu\to e\gamma$ and the branching ratio of $\mu\to 3 e$ (left) and the $\mu-e$ conversion rate within gold atoms (right) obtained from the DNN analysis in the ``T1-2B ext.'' scotogenic framework.}
    \label{Fig:LFV_T12Bext}
\end{figure}

Finally, the situation is very similar to the case of ``T1-2G'' discussed above concerning the $\mu-e$ transitions. The same correlations are observed between the different observables, and a large part of parameter space will mainly be challenged by the upcoming MEG II and COMET\footnote{COMET: search for conversion in Al, not shown in Fig.\ \ref{Fig:LFV_T12Bext}.} experiments.

\subsection{Reference scenarios based on DNN analyses}
\label{Sec:ExampleScenarios}

We finally define reference scenarios featuring fermionic dark matter in each of the two considered scotogenic frameworks. The choice is based on the highest ``DNN score'' $S$ defined in Eq.\ \eqref{Eq:CompTools:Score_function_ratio}. The resulting scenarios are listed in Table \ref{Tab:CombinedScenarios} for the ``T1-2B ext.'' and the ``T1-2G'' frameworks.

\begin{table}
    \centering
    \begin{tabular}{|c|| 
                    c@{\;\,}|c@{\;\,}|c@{\;\,}|c@{\;\,}|}
        \hline
        & \multicolumn{2}{c|}{\bf ``T1-2G''} & \multicolumn{1}{c|}{\bf ``T1-2B\,ext.''} \\
        \cline{2-4}
          & ~~Scenario I~~
          & ~~Scenario II~~  
          & ~~Scenario III~~  \\
          & {\scriptsize (doubl.\ ferm.)} 
          & {\scriptsize (tripl.\ ferm.)}  
          & {\scriptsize (doubl.\ ferm.)}  \\
        \hline\hline
        $m_{\phi^0_1}$ [GeV]        & 1862.92   & 3316.89   & 1675.12      \\
        $m_{A^0}$ [GeV]             & 2835.24   & 3981.61   & 3755.33      \\
        $m_{\phi^{\pm}_1}$ [GeV]    & 2825.56   & 3984.62   & 1675.10      \\
        \hline
        $m_{\chi^0_1}$ [GeV]        & 1047.89   & 2309.61   & 1067.62       \\
        $m_{\chi^{\pm}_1}$ [GeV]    & 1049.41   & 2309.85   & 1068.83       \\
        \hline\hline
        $m_{\mathrm{DM}}$ [GeV]     & 1047.89   & 2309.61   & 1067.62       \\
        $\Delta m\,[\mathrm{GeV}]$  & 1.52      & 0.24      & 1.21           \\
        \hline\hline
        $\sigma_{\mathrm{SI}}\,[\mathrm{cm}^2]$ 
                                    & $1.9\cdot 10^{-46}$ 
                                    & $1.6\cdot 10^{-48}$ 
                                    & $1.7\cdot 10^{-47}$ \\
        \hline
        BR($\mu\to e\gamma$)       & $2.3\cdot 10^{-13}$ 
                                   & $2.2\cdot 10^{-14}$ 
                                   & $5.0\cdot 10^{-13}$ \\
        \hline\hline
        $\chi^0_1\chi^0_1 \to VV$           & 4\%   & 9\%   & 4\%     \\
        $\chi^0_1\chi^0_2 \to q\bar q$      & 12\%  & —   & 12\%    \\
        $\chi^0_1\chi^{\pm}_1 \to q\bar q$  & 26\%  & 36\%  & 27\%    \\
        $\chi^0_1\chi^{\pm}_1 \to \ell\nu$  & 5\%   & 7\%   & 4\%    \\
        $\chi^0_1\chi^{\pm}_1 \to hW^{\pm}$ & 2\%   & —   & —   \\
        $\chi^0_2\chi^{\pm}_1 \to q\bar q$  & 18\%  & —   & 18\%   \\
        $\chi^0_2\chi^{\pm}_1 \to \ell\nu$  & 3\%   & —   & 3\%     \\
        $\chi_1^{\pm}\chi_1^{\mp} \to q\bar q$ 
                                            & 12\%  & 18\%  & 15\%    \\
        $\chi_1^{\pm}\chi_1^{\mp} \to \ell\bar\ell$ 
                                            & 3\%   & —   & 3\%    \\
        $\chi_1^{\pm}\chi_1^{\mp} \to VV$   & —   & 14\%  & —    \\
        $\chi_1^{\pm}\chi_1^{\pm} \to W^{\pm}W^{\pm}$ 
                                            & —   & 8\%   & —     \\
        \hline
    \end{tabular}
    \caption{Masses of lightest neutral and charged scalars and fermions, DM mass ($m_{\mathrm{DM}}$), mass difference between the DM candidate and the lightest charged particle ($\Delta m = m^{\pm} - m_{\mathrm{DM}}$), branching ratio of the $\mu \to e \gamma$ decay, spin-independent DM direct detection cross-section ($\sigma_{\mathrm{SI}}$), and relative contributions of the dominating DM (co-) annihilation channels for the three reference scenarios in the ``T1-2G'' and ``T1-2B ext.'' models. The channels are regrouped into classes according to $V \in \{\gamma, Z^0, W^{\pm} \}$, $q \in \{u, d, c, s, t, b \}$, $\ell \in \{e, \mu, \tau \}$ and $\nu \in \{\nu_e, \nu_{\mu}, \nu_{\tau} \}$ “—” denotes a contribution $<1\%$. The three scenarios met the relic density and Higgs mass constraints.}
    \label{Tab:CombinedScenarios}
\end{table}

As can be seen, in both cases, the scenarios are characterized by important co-annihi\-lations of the lightest neutral fermion $\chi^0_1$ with the other fermions (neutral and charged) either of the fermionic doublet or triplet, as well as co-annihilations of the other states within the doublet or triplet. This is analogous to the results presented in Refs.\ \cite{Sarazin:2021nwo, Alvarez:2023dzz, deNoyers:2024qjz} based on the MCMC analyses, where the relic density constraint has been met due to efficient co-annihilations. In the case of the singlet-like dark matter (Scenario II of ``T1-2B''), co-annihilation also are numerically dominant, as the doublet is very close in mass to the singlet.

In the same spirit as the reference scenarios proposed in Ref.\ \cite{Sarazin:2021nwo} for the ``T1-2A'' framework, and in Ref.\ \cite{deNoyers:2024qjz} for the ``T1-2G'' based on the MCMC scan, these scenarios may serve as references for future studies, e.g.\ concerning collider signatures. For practical purposes, we attach the four parameter points in SLHA \cite{SLHA1, SLHA2} format as ancillary files to the electronic submission of this paper. Finally, note that we have chosen the four scenarios with values of BR($\mu\to e\gamma$) and $\sigma_{\rm SI}$ relatively close to the current limits by MEG and LZ, so that the suggested scenarios are in the parameter range to be challenged in a near future.
\section{Performances and Perspectives}
\label{Sec:Comparison}

Comparing the obtained results, e.g.\ in terms of the dark matter mass distribution, between the two applied methods, i.e.\ the Markov Chain Monte Carlo (MCMC) method and the Deep Neural Network (DNN) analysis, it becomes clear that the two methods do not yield identical results. This being said, it is important to underline that the obtained results are \emph{per se} not incompatible, as the two techniques are not meant to give identical results. The MCMC computes the ``goodness'' of a parameter point using the likelihood. In this way, it removes evades points that are not compatible with the imposed constraints, and the distributions reflect the ``goodness'' of the individual points, thus converging towards ``best-fit'' regions of the parameter space. In contrast, the DNN identifies ``good'' and ``bad'' parameter points in view of the imposed constraints. Thus, it maps out the exclusion boundaries within the multi-dimensional parameter space, rather than searching for ``increasingly better'' points. This can, e.g., be seen in Fig.\ \ref{Fig:Histo_DM_T12G}, where the mass distribution of doublet-dominated dark matter is relatively flat between two rather well identified limits ($750 \leq m_{\rm DM} \leq 1350$ GeV), while the corresponding distribution obtained from the MCMC study peaks at $m_{\rm DM} \approx 1100 GeV$ (see Ref.\ \cite{deNoyers:2024qjz}).

While the metric for the DNN is different, we use a geometric mean here instead of a likelihood for scoring, points passing not much constraints could pass and still be classified as a valid point in a good score compared to what the likelihood would have classified it. 

The advantage of the DNN method resides in its efficiency. Indeed, the generation of points is very fast since it is purely based on randomness. Moreover, the scoring to find the points near ``good'' points is also fast. This has been optimised by using the \texttt{Tensorflow} \cite{tensorflow2015-whitepaper} library to maximise the efficiency of the workflow. The bottleneck of the used setup resides in writing the \texttt{SLHA} file for points chosen to pass by the Oracle. Indeed writing the set of parameters into \texttt{SLHA} file and give them to \texttt{SPheno}  and \texttt{MicrOMEGAs} turns out to the slowest step. A first optimisation could be to implement these two numerical codes directly into a \texttt{Python} code and thus avoid the writing of \texttt{SLHA} files. Otherwise, one could re-write the main calculation of interest to compute the spectrum and the observable to gain computation time leveraging \texttt{Python} libraries, e.g. Numpy and Scipy. 

Looking into the efficiency of the DNN, we can check how many sets of parameters were given to the Oracle. On average, within the ``T1-2G'' model, for 500 steps of the model, 225 points were considered as ``good''. At each step, we give 1000 sets of parameters to the Oracle. At the end of the computation, a final dataset of 27\,000 points considered as ``good''. This brings the number of points seen by the Oracle to 60\,000\,000 in 13 weeks, yielding an efficiency of 0.045\%. The MCMC method provides 40\,000\,000 parameter points to \texttt{SPheno}and \texttt{MicrOMEGAs} in 8 weeks, out of which 117\,000 were selected, yielding an efficiency of 0.29\%.

In conclusion, the MCMC method gives a better ratio of accepted points over tested points. Moreover, it requires less computation time. However, let us note that the major improvement of the DNN/AL method is its ability to test way more points and always scan new zones in parameter phase space. A simple example is shown in App.\ \ref{App:ToyModel} demonstrating that the AL method is capable to map the exclusion frontier defined by the Oracle. Also, if we do a fully random scan while applying the same constraints that were defined for the Oracle we estimate the amounts of ``good'' points to be close to zero within the same computation time. This shows that the DNN method brings an improvement over a full random scan, while still being based on random search. 

One way of improving this method and significantly improving the random generation could be to implement a Generative Adversarial Network (GAN), a combination of two neural networks. The first, the ``adversarial'', has to discriminate a parameter, which in the present case would be the ``goodness'' of a parameter points, i.e.\  the actual DNN developed for the present study. The second, the ``generator'' has to generate data, which in the present contect should be parameter points. The specificity of the GAN is the fact of having two DNNs training together and in a complementary way. The ``generator'' will try to ``fool'' the ``adversarial'', while it will learn itself not to guess wrongly. Combining a generator with our trained discriminator could improve the generation of parameter points, that for now rely solely on random generation. After an appropriate training period, we may thus obtain one neural network, that is efficient in discriminating parameter sets, and -- in an optimistic view -- replace {\tt SPheno} and {\tt micrOMEGAs} as it would be trained to map out the parameter space of the model used in the training. Such a setup would allow to generate important datasets on a very short timescale (a few seconds), and thus improving the scanning of the parameter space. Such an algorithm may also be useful to bypass the Casas-Ibarra parametrization and provide viable results for the neutrino sector. This may, e.g., allow to treat other models such as the scotogenic ``T1-2A ext.'' framework \cite{Alvarez:2023dzz}, which did not yield any viable results using the described AL method.
\section{Conclusion}
\label{Sec:Conclusion}

Scotogenic models are simple extensions of the Standard Model, which offer with a minimum of particles, numerous solutions to the current open questions. Indeed, the simplest extension from Ma model generates a radiative mass to neutrinos while providing a viable dark matter candidate. We saw that the models studied, the ``T1-2G'' and ``T1-2B'', feature interesting phenomenology in terms of DM and LFV. The models can provide a viable WIMP candidate, that can satisfy the relic density constraint as well as the direct detection constraints. A vast majority of the valid points showcases co-annihilation processes that help to achieve the correct relic density. In such scenarios, the mass difference between the WIMP and the next-to-lightest particle is small, typically of the order of a few $\mathrm{GeV}$. This has important implications for collider searches, as it leads to signatures related to long-lived particle. Regarding the direct detection constraints, the XENONnT experiment \cite{XENON:2020kmp} will be able to probe challenge part of the currently allowed parameter space, even though a significant fraction of the points will remain out of reach for direct detection experiments.

In terms of lepton flavour violation, the models can also provide interesting signatures. The most constraining process is the $\mu \to e \gamma$ decay. The current limit with MEG \cite{MEG:2016leq} allows points that will be challenged by the MEG II \cite{Meucci:2022qbh} experiment. New limits on the Conversion Rate of the $\mu-e$ conversion in Al nuclei will be able to probe a large part of the parameter space \cite{Haxton:2022piv}.

The Active Learning technique used in this study turned out to be more efficient than plain random sampling, but less efficient than a Markov Chain Monte Carlo setup. In future analyses, the AL technique could be upgraded to GAN algorithm in order to generate points more efficiently and then check their validity using {\tt SPheno} and {\tt micrOMEGAs}.

\section*{Acknowledgements}
The authors would like to thank W.~Porod and M.~Sarazin for useful discussions, particularly concerning the use of {\tt SARAH} and {\tt SPheno}. This work has been done thanks to the facilities offered by the Univ.\ Savoie Mont Blanc -- CNRS/IN2P3 MUST computing center. The authors would also like to thank J.~Multigner for his help to deploy the algorithm. The plots presented in this article have been obtained using {\tt MatPlotLib} \cite{MatPlotLib}. 

\paragraph{Funding information}
The work of U.\,d.\,N.\ is funded by a Ph.D.\ grant of the French Ministry for Education and Research, and M.\,D.\ is funded by a Ph.D.\ grant from the Junior Professor Chair Physique-2-DEMAIN at Univ.\ Savoie Mont Blanc with contributions from the {\it Agence Nationale de la Recherche} and the Grand Annecy. This work is supported by {\it Investissements d’avenir}, Labex ENIGMASS, contrat ANR-11-LABX-0012.

\begin{appendix}
\numberwithin{equation}{section}

\section{Details on the ``T1-2G'' and ``T1-2B ext.'' scotogenic models}
\label{App:T12B}

In this Appendix, we provide complementary information about the two scotogenic frameworks under consideration. Starting with the scalar sector, after electroweak symmetry breaking, the scalar fields are written in component notation as
\begin{equation}
    H ~=~ \begin{pmatrix} G^+ \\ \frac{1}{\sqrt{2}} \left( v + h^0 + i G^0 \right) \end{pmatrix}, ~~~~~
    \eta ~=~ \begin{pmatrix} \eta^+ \\ \frac{1}{\sqrt{2}} \left( \eta^0 + i A^0 \right) \end{pmatrix}, ~~~~~
    \Delta ~=~ \begin{pmatrix} \Delta^0/\sqrt{2} & \Delta^+ \\ \Delta^- & -\Delta^0/\sqrt{2} \end{pmatrix} \,.
    \label{Eq:Scalar_fields}
\end{equation}
Note that only the Higgs doublet $H$ acquires a vacuum expectation value, while the additional doublet $\eta$ and triplet $\Delta$ do not and hence the $\mathbb{Z}_2$ symmetry remains unbroken. Above, $G^+$, $G^0$, $h^0$, $\eta^+$, $\eta^0$ and $A^0$ are the Goldstone bosons, the physical Higgs boson, the charged scalar, the $CP$-even scalar, and the $CP$-odd pseudoscalar, respectively. The triplet $\Delta$ features charged components $\Delta^{\pm}$ and a neutral $CP$-even component $\Delta^0$.

Minimising the scalar potential, the parameter $M^2_H$ can be eliminated in favour of the Higgs self-coupling $\lambda_H$. The latter's value, at the tree-level, is about $\lambda_H \approx 0.13$.

The fermion sector of the ``T1-2B ext.'' model is composed of the SM fermions, a Dirac fermion doublet $\Psi$, and two Majorana singlets $F_k$ with $k = 1,2$. As stated in ``T1-2G'' the doublet $\Psi$ is decomposed into two Weyl fermions $\Psi_1$ and $\Psi_2$ of opposite hypercharges. The Lagrangian of the fermion sector of the ``T1-2B ext.'' model is given by
\begin{equation}
    - \mathcal{L}_{\text{fermion}} \supset M_{\Psi} \Psi_1 \Psi_2 + \frac{1}{2} M_{F_{ij}} \bar{F}_i F_j + y_1^j \Psi_1 F_j H + y_2^j \Psi_2 F_j \tilde{H} + \text{h.c.} \,,
    \label{Eq:T12B_Fermion_Lagrangian}
\end{equation}
where the sum over indices $i$ and $j$ is implicit. In this model we can also place ourselves in a basis where $M_{F_{12}} = M_{F_{21}} = 0$ without a loss of generality. Additionally, singlets $F_j$ will mix with Weyl doublets through the Yukawa couplings $y_{1j}$ and $y_{2j}$, which are supposed to be real.

In the ``T1-2B ext.'' model, after electroweak symmetry breaking, the neutral scalar states $\Delta^0$ and $\eta^0$ mix to form two $CP$-even neutral mass eigenstates $\phi_1^0$ and $\phi_2^0$. As all scalar couplings are taken to be real, no mixing occurs with the CP-odd field $A^0$. In the basis $\left\{ \Delta^0, \eta^0, A^0 \right\}$ the tree-level mass matrix of the neutral scalar sector is given by
\begin{equation}
    \mathcal{M}_{\phi^0}^{2} ~=~ \begin{pmatrix}
        M^2_{\Delta} + \frac{1}{2} v^2 \lambda_{\Delta} & - \frac{v \kappa}{2} & 0 \\
        - \frac{v \kappa}{2} & M^2_{\eta} + \frac{1}{2} v^2 \lambda_{L} & 0 \\
        0 & 0 & M^2_{\eta} + \frac{1}{2} v^2 \lambda_{A}
    \end{pmatrix},
    \label{Eq:Neutral_Scalar_Mass_Matrix_Delta_eta}
\end{equation}
where 
\begin{equation}
    \lambda_{L,A} ~=~ \lambda_{\eta} + \lambda_{\eta}^{\prime} \pm \lambda_{\eta}^{\prime\prime}\; .
    \label{Eq:Neutral_Scalar_couplings_L_A}
\end{equation}
Similar mixing occurs between the charged components of the doublet and the triplet. At tree-level, the associated mass matrix is given by
\begin{equation}
    \mathcal{M}_{\phi^{\pm}}^{2} = \begin{pmatrix}
        M_{\Delta}^2 + \frac{1}{2} v^2 \lambda_{\Delta} & \frac{v \kappa}{\sqrt{2}} \\
        \frac{v \kappa}{\sqrt{2}} & M_{\eta}^2 + \frac{1}{2} v^2 \lambda_{\eta}
    \end{pmatrix}\; .
    \label{Eq:Charged_Scalar_Mass_Delta_eta}
\end{equation}
The rotation from the interaction basis to the physical mass basis is given by the  matrices $U_{\phi^0}$ for the neutral states and $U_{\phi^{\pm}}$ for the charged ones, such that
\begin{equation}
    \begin{pmatrix}
        \phi_1^0 \\ \phi_2^0 \\ A^0
    \end{pmatrix} = U_{\phi^0} \begin{pmatrix}
        \Delta^0 \\ \eta^0 \\ A^0
    \end{pmatrix}
    \quad\text{and}\quad
    \begin{pmatrix}
        \phi_1^{\pm} \\ \phi_2^{\pm}
    \end{pmatrix} = U_{\phi^{\pm}} \begin{pmatrix}
        \Delta^{\pm} \\ \eta^{\pm}
    \end{pmatrix}\; .
    \label{Eq:Charged_Scalar_Mixing_Matrix_Delta_eta}
\end{equation}
where the $CP$-even mass eigenstates $\phi_{1,2}^{0,\pm}$ are sorted such that $\phi_{1}^{0,\pm} \leq \phi_{2}^{0,\pm}$.

In the fermion sector, and still within the ``T1-2B ext.'' framework, the neutral components of the doublets and the singlets will mix, resulting in four neutral Majorana mass eigenstates denoted by $\chi_i^0$, $i = 1,2,3,4$. In the basis $\left\{ F_1, F_2, \Psi_1^0, \Psi_2^0 \right\}$ the tree-level mass matrix of the neutral fermion sector is given by
\begin{equation}
    \mathcal{M}_{\chi^0} = \begin{pmatrix}
        M_{F_{11}} & 0 & \frac{v y_{11}}{\sqrt{2}} & \frac{v y_{21}}{\sqrt{2}} \\
        0 & M_{F_{22}} & \frac{v y_{12}}{\sqrt{2}} & \frac{v y_{22}}{\sqrt{2}} \\
        \frac{v y_{11}}{\sqrt{2}} & \frac{v y_{12}}{\sqrt{2}} & 0 & M_{\Psi} \\
        \frac{v y_{21}}{\sqrt{2}} & \frac{v y_{22}}{\sqrt{2}} & M_{\Psi} & 0
    \end{pmatrix},
    \label{Eq:Neutral_Fermion_Mass_Matrix_Psi_F}
\end{equation}
where we have set $M_{F_{12}} = M_{F_{21}} = 0$ without loss of generality. The rotation from the interaction basis to the physical mass basis is given by a matrix $U_{\chi^0}$, such that
\begin{equation}
    \begin{pmatrix}
        \chi_1^0 \\ \chi_2^0 \\ \chi_3^0 \\ \chi_4^0
    \end{pmatrix} = U_{\chi^0} \begin{pmatrix}
        F_1 \\ F_2 \\ \Psi_1^0 \\ \Psi_2^0
    \end{pmatrix}\; ,
    \label{Eq:Neutral_Fermion_Mixing_Matrix_Psi_F}
\end{equation}
where the mass eigenstates are ordered such that $m_{\chi_i^0} \leq m_{\chi_j^0}$ for $i,j = 1,2,3,4$. The charged fermion mass is already given by $M_{\Psi}$ so
\begin{equation}
    m_{\chi^{\pm}} = M_{\Psi}\; .
    \label{Eq:Charged_Fermion_Mass_Psi_F}
\end{equation}

For the ``T1-2G'' framework, a detailed discussion of scalar and fermion mixing can be found in Ref.\ \cite{deNoyers:2024qjz}.

\section{Conditions on scalar couplings parameters}
\label{App:ScalarConditions}

To insure vacuum stability, the potential has to be bounded from below. This translates with the following conditions on the couplings for the ``T1-2G'' and ``T1-2A ext.'' models
\begin{equation}
    \begin{split}
        2 \lambda_H ~&>~ \frac{M^2_H}{M^2_{\eta}} \left(\lambda_{\eta} + \lambda^{\prime}_{\eta} + \lambda^{\prime \prime}_{\eta}\right) \,, \\
        2 \lambda_H M^2_S ~&>~ \lambda_S M^2_H \,, \\
        \lambda_{\eta} ~&>~ -2 \sqrt{\lambda_H \lambda_{4 \eta}} \,, \\
        \lambda_{\eta} + \lambda^{\prime}_{\eta} - \big|\lambda^{\prime \prime}_{\eta}\big| ~&>~ -2 \sqrt{\lambda_H \lambda} \,, \\
        \frac{1}{2} \lambda_S ~&>~ -2 \sqrt{\lambda_H \lambda_{4S}} \,, \\
        \frac{1}{2} \lambda_{S\eta} ~&>~ -2 \sqrt{\lambda_{4\eta} \lambda_{4S}} \,.
    \end{split}
    \label{Eq:BoundednessFromBelow1_T12G_T12A}
\end{equation}
The scalar couplings also have to satisfy:
\begin{equation}
    \begin{split}
        \sqrt{8 \lambda_H \lambda_{4\eta} \lambda_{4S}} + \lambda_{\eta} \sqrt{2 \lambda_{4S}} + \frac{1}{2} \lambda_{S} \sqrt{2 \lambda_{4 \eta}} + \frac{1}{2} \lambda_{S \eta} \sqrt{2 \lambda_H} + \rho_1 ~&>~ 0 \\
        \sqrt{8 \lambda_H \lambda_{4\eta} \lambda_{4S}} + \left(\lambda_{\eta} + \lambda^{\prime}_{\eta} - \big|\lambda^{\prime \prime}_{\eta}\big|\right) \sqrt{2 \lambda_{4S}} + \frac{1}{2} \lambda_{S} \sqrt{2 \lambda_{4 \eta}} + \frac{1}{2} \lambda_{S \eta} \sqrt{2 \lambda_H} + \rho_2 ~&>~ 0 \, ,
    \end{split}
    \label{Eq:BoundednessFromBelow2_T12G_T12A}
\end{equation}
with $\rho_{1,2}$ being defined as 
\begin{equation}
    \begin{split}
        \rho_1 ~&=~ \sqrt{2 \left(\lambda_{\eta} + 2 \sqrt{\lambda_H \lambda_{4 \eta}}\right)\left(\frac{1}{2}\lambda_{S} + 2 \sqrt{\lambda_H \lambda_{4S}}\right)} \\
        \rho_2 ~&=~ \sqrt{2 \left(\lambda_{\eta} + \lambda^{'}_{\eta} - \big|\lambda^{''}_{\eta}\big| + 2 \sqrt{\lambda_H \lambda_{4 \eta}}\right)\left(\frac{1}{2}\lambda_{S} + 2 \sqrt{\lambda_H \lambda_{4S}}\right)} \, .
    \end{split}
    \label{Eq:CoefficientsRho_T12G_T12A}
\end{equation}

For the ``T1-2B ext.'' model, this requirement of vacuum stability translates into the following conditions on the couplings
\begin{equation}
    \begin{split}
        \lambda_{4 \eta} ~&\geq~ 0 \,, \\
        \lambda_{4 \Delta} ~&\geq~ 0 \,, \\
        \lambda_{\Delta} &\leq \left| 2 \sqrt{3 \lambda_{4\Delta} \lambda_H} \right| \,, \\
        \lambda_{\phi} &\leq \left| \sqrt{2 \lambda_{4\phi} \lambda_H} \right| \,, \\
        \lambda_{\Delta \phi} &\leq \left| 2 \sqrt{6 \lambda_{4\phi} \lambda_{4\Delta}} \right| \,. 
    \end{split}
    \label{Eq:BoundednessFromBelow_T12B}
\end{equation}

\section{\texorpdfstring{Expression of the loop matrix $M_L$}{Expression of the loop matrix M\_L}}
\label{App:LoopMatrix}

The elements of the $3 \times 3$ matrix $M_L$ appearing in Eq.\ \eqref{Eq:NeutrinoMassMatrix}, expressed in terms of the scalar and fermionic masses and mixing matrices, read
\begin{equation}
    \begin{split}
    \left( M_L \right)_{11} &= \sum_{k,n} b_{kn}  \, (U_{\chi}^{\dagger})^2_{4k} \, (U_{\phi}^{\dagger})^2_{1n} \,, \\
    \left( M_L \right)_{22} &= \frac{1}{2} \sum_{k,n} b_{kn} \, (U_{\chi}^{\dagger})^2_{1k} \, \Big[ (U_{\phi}^{\dagger})^2_{2n} - \big( U_{\phi}^{\dagger} \big)^2_{3n} \Big]  \,, \\
    \left( M_L \right)_{33} &= \frac{1}{2} \sum_{k,n} b_{kn}  \, (U_{\chi}^{\dagger})^2_{2k} \, \Big[ (U_{\phi}^{\dagger})^2_{2n} - \big( U_{\phi}^{\dagger} \big)^2_{3n} \Big] \,, \\
    \left( M_L \right)_{12} = \left( M_L \right)_{21} &= \frac{1}{\sqrt{2}}\sum_{k,n} b_{kn}  \, (U_{\chi}^{\dagger})_{1k} \, (U_{\chi}^{\dagger})_{4k} \, (U_{\phi}^{\dagger})_{1n} \, (U_{\phi}^{\dagger})_{2n} \,, \\
    \left( M_L \right)_{13} = \left( M_L \right)_{31} &= \frac{1}{\sqrt{2}} \sum_{k,n} b_{kn} \, (U_{\chi}^{\dagger})_{2k} \, (U_{\chi}^{\dagger})_{4k} \, (U_{\phi}^{\dagger})_{1n} \, (U_{\phi}^{\dagger})_{2n}  \,, \\
    \left( M_L \right)_{23} = \left( M_L \right)_{32} &= \frac{1}{2} \sum_{k,n} b_{kn}  \, (U_{\chi}^{\dagger})_{1k} (U_{\chi}^{\dagger})_{2k} \, \Big[ (U_{\phi}^{\dagger})^2_{2n} - \big( U_{\phi}^{\dagger} \big)^2_{3n} \Big] \,, \\
    \end{split}
    \label{Eq:LoopComponents}
\end{equation}
where the sums run over the neutral fermion mass eigenstates $\left(k= 1,2,3,4\right)$ and the neutral scalar mass eigenstates $\left(n = 1,2,3, \; \text{the last one corresponding to the pseudo-scalar $A^0$} \right) $. The coefficients $b_{kn}$, stemming from the loop integrals, are given by 
\begin{equation}
    b_{kn} = \frac{1}{16 \pi^2} \frac{m_{\chi^0_k}}{m^2_{\phi^0_n} - m^2_{\chi^0_k}} \left[ m^2_{\chi^0_k} \ln{\left(m^2_{\chi^0_k}\right)} - m^2_{\phi^0_n} \ln{\left(m^2_{\phi^0_n}\right)}\right] \,.
    \label{Eq:LoopCoefficients}
\end{equation}
We note that there is no dependence on the renormalisation scale, as the terms given in Eq.\ \eqref{Eq:LoopComponents} constitute the leading order contribution.

\section{Matrices involved in the Casas-Ibarra parametrization}
\label{App:CasasIbarra}

The PMNS matrix takes the following shape
\begin{equation}
    U_{\text{PMNS}} ~=~ \!\! \begin{pmatrix} 
		    1 & 0 & 0 \\ 0 & c_{23} & s_{23} \\ 0 & -s_{23} & c_{23} 
		\end{pmatrix} \!\!
		\begin{pmatrix} 
		    c_{13} & 0 & s_{13} e^{-i \delta_{\rm CP}} \\ 
		    0 & 1 & 0 \\ 
		    -s_{13}e^{i \delta_{\rm CP}} & 0 & c_{13} 
		\end{pmatrix} \!\!
		\begin{pmatrix} 
		    c_{12} & s_{12} & 0 \\ -s_{12} & c_{12} & 0 \\ 0 & 0 & 1 
		\end{pmatrix} \!\!
		\begin{pmatrix} 
		    1 & 0 & 0 \\ 
		    0 & e^{i \alpha_{1}} & 0 \\ 
		    0 & 0 & e^{i \alpha_{2}}
        \end{pmatrix} \,,
    \label{Eq:PMNSMatrix}
\end{equation}
with $\delta_{\rm CP}$ taking its value in the range from Ref.\ \cite{NuFit2020}, and the Majorana phases $\alpha$ being between $0$ and $\pi$.

The rotation matrix $R$ appearing in Eq.\ \eqref{Eq:CasasIbarra} can be written as
\begin{equation} 
     R ~=~ \left( \begin{array}{ccc}
         \sqrt{1-r_1^2 } & -r_1 & 0  \\
         r_1 & \sqrt{1-r_1^2 } & 0 \\
         0 & 0 & 1
     \end{array} \right) \left( \begin{array}{ccc}
         \sqrt{1-r_2^2 } & 0 & r_2  \\
         0 & 1 & 0 \\
         -r_2 & 0 & \sqrt{1-r_2^2 }
     \end{array} \right) \left( \begin{array}{ccc}
         1 & 0 & 0  \\
         0 & \sqrt{1-r_3^2 } & -r_3 \\
         0 & r_3 & \sqrt{1-r_3^2 }
     \end{array} \right) \; ,
     \label{eq:Rmat}
\end{equation}
depending on three complex parameters $r_k$ ($k=1,2,3$) defined in the Sec. Setup.

\section{Toy model illustrating the Deep Neural Network approach}
\label{App:ToyModel}

In order to get a better understanding on the Deep Neural Network (DNN) and the Active Learning (AL) algorithm, we have tested the chosen realization on a simple toy model. The idea is to discriminate two parameters $x$ and $y$ in a two-dimensional space. The parameters are both chosen in the range $[0,1]$ and the points are classified as ``good'' (${\cal C}=1$) or ``bad'' (${\cal C}=0$) according to whether or not they fall inside or outside an ellipse parametrized through the condition
\begin{equation}
    \mathcal{C} = \left\{
    \begin{array}{ll}
        1 & \text{~~~~if } \big(x-0.2\big)^2 + \big(y-0.6\big)^2 \leq 0.25\\
        0 & \text{~~~~otherwise}.
    \end{array}
    \right.
    \label{Eq:CompTools:Toy_model_condition}
\end{equation}
In this toy model, we chose the settings:
\begin{itemize}
    \item $K=100$, number of set of parameter given to the oracle;
    \item $L=5000$, size of set of parameter to choose the good one near from each other;
    \item $\alpha=5\cdot 10^{-3}$, diversity weighting;
    \item $N=200$, number of iterations.
\end{itemize}

\begin{figure}
    \centering
    \includegraphics[width=0.6\textwidth]{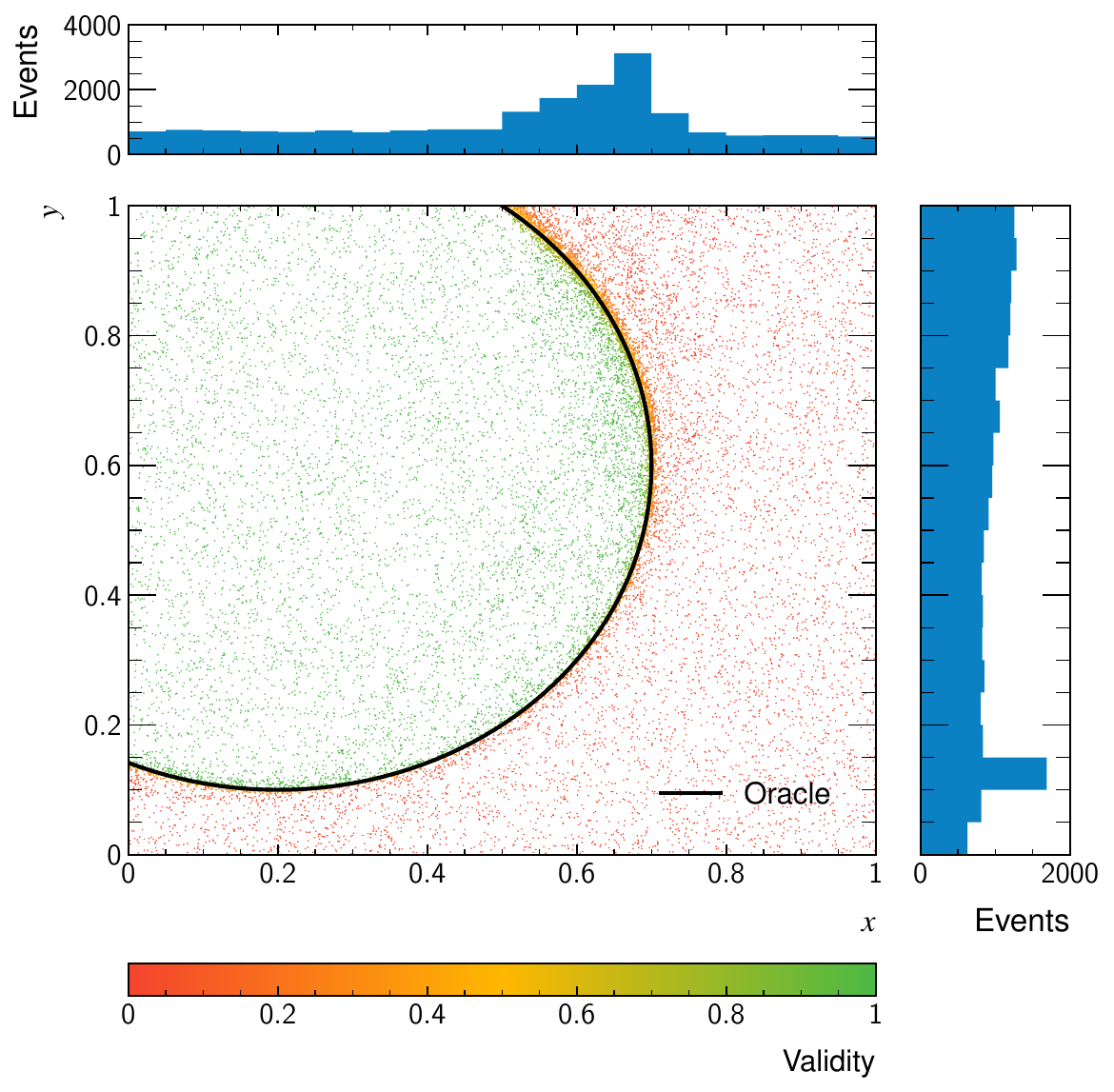}
    \caption{Toy model illustration of the DNN and the AL algorithm. The resulting ``good'' and ``bad'' points are represented in green and red, respectively. The black line is the decision boundary between the two cases. The blue histograms indicate the density of points projected in the two axes.}
    \label{Fig:Toy_model}
\end{figure}

The obtained results are shown in Fig.\ \ref{Fig:Toy_model}. As can be seen, the DNN is able to learn the decision boundary between the two classes. The density of points, depicted in the histograms, is maximized around this boundary. This two-dimensional toy model is a good test for the DNN and the AL algorithm. Mapping out the exclusion boundary between the phenomenologically viable (``good'') and non-viable (``bad'') points in a high-dimensional model proceeds in the same way, e.g.\ for the scotogenic models considered in this paper. It is, however, complicated, if not impossible, to produce a graphical representation of such high-dimensional parameter spaces.
\end{appendix}

\bibliography{Refs/Refs}

\end{document}